\documentclass[useAMS]{aaifm}

\usepackage{amssymb}
\usepackage{mathrsfs}
\usepackage{graphicx}

\newcommand\Ai{\mbox{Ai}}            

\newcommand{\df}[2]{\displaystyle\frac{#1}{#2}}
\newcommand{\tf}[2]{\textstyle\frac{#1}{#2}}

\newcommand{\Int}[2]{\displaystyle\int_{#1}^{#2}}

\newcommand{\PDD}[2]{\df{\partial #1}{\partial #2}}
\newcommand{\PD}[1]{\df{\partial }{\partial #1}}

\newcommand{\PDDT}[2]{\df{\partial^{2} #1}{\partial #2^{2}}}
\newcommand{\ODDT}[2]{\df{{\rm d}^{2} #1}{{\rm d} #2^{2}}}

\newcommand{\ub}[2]{\underbrace{#1}_{#2}}

\newcommand{\ts}[1]{\tilde{#1}}

\newcommand{\eps}{\varepsilon}
\newcommand{\B}[1]{\mbox{\boldmath $ #1 $} }

\newcommand{\D}[1]{{#1}^{\prime}}
\newcommand{\DD}[1]{{#1}^{\prime\prime}}
\newcommand{\DDD}[1]{{#1}^{\prime\prime\prime}}

\newcommand{\md}{|}
\newcommand{\be}{\begin{eqnarray}}
\newcommand{\en}{\end{eqnarray}}
\newcommand{\no}{\nonumber}

\newcommand{\half}{\tiny\frac{1}{2}}
\newcommand{\thalf}{\tiny\frac{3}{2}}
\newcommand{\fhalf}{\tiny\frac{5}{2}}

\newcommand{\beqa}{\begin{eqnarray}}
\newcommand{\eeqa}{\end{eqnarray}}

\begin{document}

\title[Interaction of Tollmien-Schlichting Waves with the Sea Surface]
{Interaction of Tollmien-Schlichting Waves in the Air with the Sea
Surface}

\author[S.G. Sajjadi \& H. Khanal]{Shahrdad G. Sajjadi and Harihar Khanal}

\affiliation{Center for Geophysics and Planetary Physics,
Department of Mathematics, Embry-Riddle Aeronautical University,
Daytona Beach, FL, U.S.A.\\emails: Shahrdad.Sajjadi@erau.edu,
harihar.khanal@erau.edu}

\maketitle

\begin{abstract}{Tollmien-Schlichting waves, Hydrodynamic
stability, Air-sea interaction, Two-fluid interface, Solitons}
Linear stability of fully developed flows of air over water is
carried out in order to study non-linear effects in the generation
of solitons by wind. A linear stability analysis of the basic flow
is made and the conditions at which solitons first begin to grow
is determined. Then, following [10], the non-linear stability of
the flow is examined and the quintic non-linear Schr\"{o}dinger
equation is derived for the amplitude of disturbances. The
coefficients of the non-linear Schr\"odinger equation are
calculated from the eigenvalue problem which determines the
stability of air-water interface.

An asymptotic and a numerical stability analysis is carried out to
determine the neutrally stable flow conditions for air-sea
interface. Four different profiles are considered for the airflow
blowing over the surface of the sea, namely, plane Couette flow
(pCf), plane Poiseuille flow (pPf), laminar and turbulent boundary
layer (L,TBL) profiles. For each of the above cases the shear flow
counterpart in the water is assumed to be a pPf.

A nonlinear stability analysis results in the nonlinear
Schr\"odinger equation \be
\PDD{A}{\tau}-a_2\PDDT{A}{\xi}=\df{d_1}{d_{1r}}A+\kappa
A|A|^2+\varpi A|A|^4, \no \en where $\xi$ and $\tau$ are
local variables, and the amplitude of the surface wave is
proportional to $A$. The complex constants $a_2$, $d_1$, $\kappa$
and $\varpi$ are evaluated from the linear stability of the
two-fluid interface. The profile for the initial condition
considered here is that of the Stokes wave \be \eta=a\cos
kx+\tf{1}{2}a^2k\cos 2kx, \no \en
where $a$ is the amplitude and $k$ is the wavenumber of the
surface Stokes wave.

It is shown that the above amplitude equation produces `snake'
solitons [9] for pCf, pPf and LBL profiles, with striking
similarities. On the other hand, for TBL we observe a very violent
surface motion. For cases of pCf and LBL remarkable similarity is
observed with observations made at sea.

We conclude that the effect of nonlinearity in the airflow over
the sea surface is much larger than nonlinear interactions in the
water, and hence it is not possible to decouple the motion in the
air and the water for finite amplitude wind-wave interactions,
particularly in the case of wind-generated solitons in shallow
waters.
\end{abstract}

\section{Introduction}
The wind-generated solitons which moves nondispersively on the
thin wind-driven drift layer of wind-excited surface wave waves
are important because they are products of the processes under
which surface waves are generated by wind. When the local
atmospheric conditions are such that significant bubble densities
are created, principally by breaking waves, the resulting bubble
layer will dominate and mask these solitons, in underwater
acoustic scattering. A qualitative mixed linear-nonlinear model of
ocean-wave generation is briefly described, a key element of which
is the generation of solitons by wind. These solitons are created,
and destroyed on moving gravity waves, caused by waves being
generated by wind along with accompanying capillary components.
The gravity-capillary waves are dispersive, in contrast the
solitons are nondispersive, that is to say, they move at constant
speed, on the thin moving wind drift layer generated by the upper
part of the water surface.

In this paper, we will examine the stability of laminar and
turbulent flows of air over water surface where viscous effects
are taken into account in both fluids. The flow that will be
studied here is the fully developed flow of air over water. In
contrast to [2], we shall restrict our analysis to only the case
when the motion is driven solely by a constant pressure gradient.

A linear stability analysis will be performed to determine the
neutrally stable flow conditions. This approach is different to
[4], or indeed other similar theories such as [1], and most
experimental work, where the emphasis is on determining wave
growth rates rather than the conditions at which waves will first
start to grow. Following [10], a non-linear analysis will be
performed which will yield the non-linear quintic Schr\"{o}dinger
equation

\be
\PDD{A}{\tau}-a_2\PDDT{A}{\xi}=\df{d_1}{d_{1r}}A+\kappa
A|A|^2+\varpi A|A|^4, \label{1.1} \en governing the wave
amplitude. In equation (\ref{1.1}) $A$ is proportional to
amplitude of the non-linear surface wave. The scaled length and
time, $\xi$ and $\tau$, and the complex constants $a_2, d_1$ and
$\kappa$ and $\varpi$ will be defined in the proceeding analysis.
The constants $a_2$ and $d_1$ are properties of the flow which is
obtained from the linear stability theory, while the effect of
non-linear interactions is determined by $\kappa$ and $\varpi$.

It is to be noted that although equation (\ref{1.1}) is strictly
applicable to velocity profiles of laminar flow, there are
experimental and theoretical suggestions that equations similar to
that of (\ref{1.1}) are also applicable to wind-wave generation
situations. For example, the experimental work [6] suggests that
wind-waves at fixed fetch and under steady wind conditions have
properties that are similar to non-linear Stokes wavetrains.
Indeed, the wind-waves given by the equilibrium amplitude
solutions of (\ref{1.1}) (for which $\kappa_{r}<0$ and $\varpi=0$)
are similar to Stokes wavetrains \be
\eta_w=a\cos\left\{k_cx-k_cc_rt\left[\tf{1}{2}\alpha(ak_c)^2+\beta(ak_c)^2
\right]\right\}\label{1.2} \en where $k_c$ is the critical
wavenumber, $\alpha$ and $\beta$ are the dimensionless energy
transfer parameters [13], and $c$ is the wave phase speed. It has
been shown [11] that such wavetrains with non-linear effects lead
to changes in the phase speed of the wave and provides the
sideband instability mechanism, similar to that of Benjamin-Fier,
which will eventually result to the break-up of the wave-train.

The wind-generated solitons which moves nondispersively on the
thin wind-driven drift layer of wind-excited surface wave waves
are important because they are products of the processes under
which surface waves are generated by wind. When the local
atmospheric conditions are such that significant bubble densities
are created, principally by breaking waves, the resulting bubble
layer will dominate and mask these solitons, in underwater
acoustic scattering. A qualitative mixed linear-nonlinear model of
ocean-wave generation is briefly described, a key element of which
is the generation of solitons by wind. These solitons are created,
and destroyed on moving gravity waves, caused by waves being
generated by wind along with accompanying capillary components.
The gravity-capillary waves are dispersive, in contrast the
solitons are nondispersive, that is to say, they move at constant
speed, on the thin moving wind drift layer generated by the upper
part of the water surface.

In this paper, we will examine the stability of laminar and
turbulent flows of air over water surface where viscous effects
are taken into account in both fluids. The flow that will be
studied here is the fully developed flow of air over water. In
contrast to [2], we shall restrict our analysis to only the case
when the motion is driven solely by a constant pressure gradient.

\section{Formulation of the problem}

We consider the steady, flow of two super-imposed fluid layers
confined between a stationary horizontal parallel plate, placed at
$y=-D_\ell$, and open to atmosphere above at a distance $y=D_u\gg
1$. We assume the flow of the upper fluid is turbulent, satisfying
Prandtl's mixing-length model, and that of the lower one is a
fully developed laminar flow. The velocity vector
$\B{U}=[U_b(y),0,0]$, for the upper fluid, and
$\B{u}=[u_b(y),0,0]$, for the lower fluid,\footnote{Hereafter, the
upper and the lower symbols (Greek or Roman) denote the upper
(air) and the lower (water) fluids, respectively. Also the
subscripts $a$ and $w$ denote air and water, respectively.} both
satisfy the Navier-Stokes equations. In the present case we shall
confine ourselves to the case when the motion of the lower fluid
is driven solely by a constant pressure gradient. Thus, the
equations of motion and their corresponding boundary conditions
reduce to
\begin{eqnarray}
\left.\begin{array}{lll}
U_b\rightarrow U_\infty, & & \mbox{as}\hspace*{0.25cm} y\rightarrow D_u\\
l^2\left(\df{dU_b}{dy}\right)^2=U_*^2;\,\,l=\kappa(y+y_0),
& & \mbox{in upper fluid}\\
u_b=U_b,\hspace*{0.25cm}\mu_w\df{du_b}{dy}=\mu_a\df{dU_b}{dy},
& & \mbox{on}\hspace*{0.25cm} y=0\\
\df{G}{\rho_w}+\nu_w\ODDT{u_b}{y}=0, & & \mbox{in lower fluid}\\
u_b=0, & & \mbox{on}\hspace*{0.25cm} y=-D_\ell\\
\end{array}\right\}\label{2.1}
\end{eqnarray}
where
$-G=\df{dp}{dx}.$
The solutions of
(\ref{2.1}) are given by
\begin{eqnarray}
U_b(y)=U_1\ln\left(\df{y+y_0}{y_0}\right)+U_0
\label{2.3}\\
u_b(y)=\df{G}{2\mu_w}(D^2_\ell-y^2)+\df{\mu_a}{\mu_w}
\df{U_1}{y_0}(D_\ell+y)\label{2.4}
\end{eqnarray}
where \be
U_0=\df{D_\ell}{\mu_w}\left(\df{U_1\mu_a}{y_0}+\df{GD_\ell}{2}\right),
\hspace*{0.5cm}U_1=\df{U_*}{\kappa}\label{2.4a} \en $U_*$ is the
friction velocity of the upper fluid, $\kappa$ is the von
K\'arm\'an's constant, and $y_0$ is an effective roughness length.

Having obtained the velocity fields governing the basic flow, we
shall now formulate the non-dimensional equations that govern
linear instability. The motion is governed by the
three-dimensional Navier-Stokes equations in each layer of fluid.
The boundary conditions are the no-slip conditions on the plates
and continuity of stress and velocity across the unknown interface
$y=\zeta(x,z,t)$, which will be determined from the proceeding
non-linear analysis.

We will introduce the following curvilinear coordinates [5]
\begin{eqnarray}
\left. \begin{array}{l} \chi=x-\df{a\cosh k(y\mp d)}
{\sinh kd}\sin k(x-ct)\\
\\
\eta=y-\df{a\sinh k(y\mp d)}
{\sinh kd}\cos k(x-ct)\\
\\
\vartheta=z
\end{array} \right\} \label{2.6}
\end{eqnarray}
where in (\ref{2.6}), the minus sign is for the upper media and
the plus sign for the lower media. The Jacobian of the
transformation is
\begin{eqnarray}
J=1-\df{2ak}{\sinh kd}\cosh k(\eta\mp d) \cosh
k(\chi-ct)+O(ak)^2\label{2.7}
\end{eqnarray}
Note that, in the linear stability theory we shall be restricted
to infinitesimal disturbances for which the Jacobian (\ref{2.7})
reduce to
\begin{eqnarray}
J=1+O(ak)\nonumber
\end{eqnarray}
Thus, within this order of approximation, the governing equations
are symbolically the same as that of Cartesian coordinates. We
remark that under the transformation (\ref{2.6}), the lines $y=d$
and $y=-d$ will map to lines $\eta=1$ and $\eta=-d^{-1}$,
respectively, and the unknown interface $y=\zeta$ will then be at
$\eta=0$.

The boundary conditions are:
\newline on $\eta=1$ no-slip conditions
$\hspace*{1.2cm}U=V=W=0$,
\newline on $\eta=0$ (the unknown interface);
\newline (i) the continuty of velocities
$\hspace*{0.9cm} \B{u}=\B{U}$
\newline(ii) the kinematic conditions
\begin{eqnarray}
\left.\begin{array}{l}
\PDD{\zeta}{t}+U\PDD{\zeta}{\chi}+V\PDD{\zeta}{\eta}=W\\
\\
\PDD{\zeta}{t}+u\PDD{\zeta}{\chi}+v\PDD{\zeta}{\eta}=w\\
\end{array}\right\}\label{2.10}
\end{eqnarray}
(iii) the continuity of tangential stresses
\begin{eqnarray}
\left.\begin{array}{l}
\PDD{U}{\eta}+\PDD{V}{\chi}=\mu^{-1}\left(\PDD{u}{\eta}+\PDD{v}{\chi}\right)\\
\\
\PDD{W}{\eta}+\PDD{V}{\vartheta}=\mu^{-1}\left(\PDD{w}{\eta}+\PDD{v}{\vartheta}\right)\\
\\
\PDD{U}{\vartheta}+\PDD{W}{\chi}=\mu^{-1}\left(\PDD{u}{\vartheta}+\PDD{w}{\chi}\right)\\
\end{array}\right\}\label{2.11}
\end{eqnarray}
(iv) the dynamic condition
\begin{eqnarray}
-\rho P+\df{2\rho}{R}\PDD{V}{\eta}+\rho{\mathscr F}\zeta=-\rho
p+\df{2}{R\nu} \PDD{v}{\eta}-{\mathscr T}\nabla_\parallel^2\zeta+{\mathscr
F}\zeta\label{2.12}
\end{eqnarray}
where $\nabla_\parallel^2\equiv\partial^2/\partial \chi^2+
\partial^2/\partial \eta^2$, and
$d=D_u/D_\ell,\mu=\mu_a/\mu_w,
\rho=\rho_a/\rho_w,
\nu=\nu_a/\nu_w.$
Also, on $\eta=-d^{-1}$  there is no-slip conditions
$u=v=w=0$.

In (\ref{2.12}) $R=U_\infty d/\nu_a$ is the Reynolds number,
${\mathscr F}=gd/U^2_\infty$ is the Froude number and ${\mathscr
T}=S/\rho_wU_\infty^2d$ is the Weber number, where $U_\infty$ is a
reference velocity, $S$ is the dimensional surface tension and $g$
is the acceleration due to gravity.

We next perturb the basic flow by infinitesimal disturbances, i.e.
\begin{eqnarray}
& &\B{U}=[U_b(\eta),0,0]+\delta[U(\eta),V(\eta),W(\eta)]
e^{i(k\chi+\ell \vartheta-kct)}+{\rm c.c.}\nonumber\\
& &\B{u}=[u_b(\eta),0,0]+\delta[u(\eta),v(\eta),w(\eta)]
e^{i(k\chi+\ell \vartheta-kct)}+{\rm c.c.}\nonumber\\
& &P=\df{\chi\DD{U}_b}{\rho\mu R}+\delta P(\eta) e^{i(k\chi+\ell
\vartheta-kct)}+{\rm c.c.}\nonumber\\
& &p=\df{\chi\DD{u}_b}{\rho\mu R}+\delta p(\eta) e^{i(k\chi+\ell
\vartheta-kct)}+{\rm c.c.}\nonumber\\
& &\zeta=\delta a e^{i(k\chi+\ell \vartheta-kct)}+{\rm c.c.},\,\,\,\md\delta\md\ll 1\no
\end{eqnarray}
where c.c. denotes a complex conjugate, then the linearized equations of motion may be cast as
\be
\left.\begin{array}{lcl}
U=V=W=0 & & \mbox{on $\eta=1$}\\
\\
ik(U_b-c)U+VU_b'=-ikP+R^{-1}[U''-(k^2+\ell^2)U] & & \\
\\
ik(U_b-c)V=-P'+R^{-1}[V''-(k^2+\ell^2)V] & & \\
\\
ik(U_b-c)W=-i\ell P+R^{-1}[W''-(k^2+\ell^2)W] & & \\
\\
ikU+V'+i\ell W=0 & & \\
\\
u+au'_b=U+aU_b',\,\,\,v=V,\,\,w=W & & \mbox{on $\eta=0$}\\
\\
ik(u_b-c)a=v,\,\,\,ik(U_b-c)a=V & & \mbox{on $\eta=0$}\\
\\
u'+ikv=\mu(U'+ikV),\,\,\,w'+i\ell v=\mu(W'+i\ell V) & & \mbox{on $\eta=0$}\\
\\
\rho p-{\mathscr F}a-2v'/\nu R-{\mathscr T}a(k^2+\ell^2)=\rho \left(P-{\mathscr F}a-2V'/R\right) & & \mbox{on $\eta=0$}\\
\\
u=v=w=0 & & \mbox{on $\eta=d^{-1}$} \\
\\
ik(u_b-c)u+vu_b'=-ik\rho p+(\nu R)^{-1}[u''-(k^2+\ell^2)u] & & \\
\\
ik(u_b-c)v=-\rho P'+(\nu R)^{-1}[v''-(k^2+\ell^2)v] & & \\
\\
ik(u_b-c)w=-i\ell\rho P+(\nu R)^{-1}[w''-(k^2+\ell^2)w] & & \\
\\
iku+v'+i\ell w=0 & & \\
\end{array}\right\}
\en Substituting in equations of motion and their corresponding
boundary conditions, making use of the Squire transformation [14]
$m^2=k^2+\ell^2,\hspace*{0.25cm}\hat{U}=m^{-1}(kU+\ell
W),\hspace*{0.25cm}\hat{V}=V,\hspace*{0.25cm}\hat{c}=c,
\hat{P}=(m/k)P,\hspace*{0.25cm}\hat{a}=(k/m)a$, etc. (dropping the
hat notation) and introducing the stream functions $\Phi(\eta)$
and $\phi(\eta)$ such that
\begin{eqnarray}
\left.\begin{array}{lll}
U=\D{\Phi}, & & u=\D{\phi}\\
 & & \\
V=-ik\Phi, & & v=-ik\phi
\end{array}\right\}\label{2.14}
\end{eqnarray}
we obtain, after linearization and some manipulations,
\begin{eqnarray}
\left.\begin{array}{lll}
\Phi=\D{\Phi}=0, & & \mbox{on}\hspace*{0.25cm} \eta=1\\
\\
ik(U_b-c)(\DD{\Phi}-k^2\Phi)-ik\DD{U}_b\Phi= & & \\
 & &\\
\hspace*{3cm}R^{-1}(\Phi^{iv}-2k^2\DD{\Phi}+k^4\Phi) & &
\mbox{in upper fluid}\\
\\
\left.\begin{array}{l}
\phi=\Phi\\
\\
\D{\phi}-\df{\D{u}_b\phi}{\sigma}=
\D{\Phi}-\df{\D{U}_b\Phi}{\sigma}\\
\\
\DD{\phi}+k^2\phi=\mu(\DD{\Phi}+k^2\Phi)\\
\\
\df{1}{\nu R}(\DDD{\phi}-3k^2\D{\phi})+ik(\phi u_b-
\D{\phi}\sigma)+i\df{k\phi}{\sigma}\df{({\sf F}+k^2{\sf T})}{R^2}=\\
\\
\df{\rho}{R}(\DDD{\Phi}-3k^2\D{\Phi})+
\rho ik(\Phi U_b-\D{\Phi}\sigma)+\rho i\df{k\Phi}{\sigma}\df{{\sf F}}{R^2}\\
\end{array}\right] & & \mbox{on}\hspace*{0.25cm} \eta=0\\
\\
ik(u_b-c)(\DD{\phi}-k^2\phi)-ik\DD{u}_b\phi= & & \\
 & & \\
\hspace*{3cm}(\nu R)^{-1}(\phi^{iv}-2k^2\DD{\phi}+k^4\phi) & &
\mbox{in lower fluid}\\
\\
\phi=\D{\phi}=0, & & \mbox{on}\hspace*{0.25cm} \eta=-d^{-1}\\
\end{array}\right\}\label{2.15}
\end{eqnarray}
where $\sigma=u_b(0)-c=U_b(0)-c$. Note that, we have redefined the
Froude and Weber number to be ${\sf F}=gD_u^3/\nu_a^2$ and ${\sf
T}=SD_u/\rho_w\nu_a^2$, respectively.

Systems similar to that of (\ref{2.15}) have been derived
[7,15--17]. However, as was pointed out by Blennerhassett [2], in
the derivations [7,17] the relevance of ${\mathscr F}$ and ${\mathscr T}$
to the stability of two-dimensional waves was not noted. He
further remarks that in earlier derivations [15,16], for the
stability of superposed fluid layers, the terms $u_b'\phi/\sigma$
and $U_b'\Phi/\sigma$ was incorrectly omitted from the second
interface condition in the system (\ref{2.15}).

We remark that in general the solution of (\ref{2.15}), and determining the
location of the neutral stability curve in $(R,k)$--plane is a computational
task. The full details of the numerical methods is given in section 6 below.

\section{The long-wave limit}

Following [2], we consider the stability of the basic flow in the
limit of long wavelength disturbances and expand the
eigenfunctions and the eigenvalue in regular perturbation series
in $k$ [7]. Thus we may write
\begin{eqnarray}
\left.\begin{array}{l} c=c_0+kc_1+...\\
\\
\phi=\phi_0+k\phi_1+...\\
\\
\Phi=\Phi_0+k\Phi_1+...\\
\end{array}\right\}\label{2.16}
\end{eqnarray}
as $k\rightarrow 0$ for a fix value of $R, {\sf F}, {\sf T}$.
Substituting (\ref{2.16}) in (\ref{2.15}) we obtain up to $O(1)$
the following homogeneous system
\begin{eqnarray}
\left.\begin{array}{lll}
\Phi_0=\D{\Phi}_0=0, & & \mbox{on}\hspace*{0.25cm} \eta=1\\
\\
0=R^{-1}\Phi_0^{iv} & & \mbox{in upper fluid}\\
\\
\\
\left.\begin{array}{l}
\phi_0=\Phi_0\\
\\
\D{\phi}_0-\df{\D{u}_b\phi_0}{\sigma}=
\D{\Phi}_0-\df{\D{U}_b\Phi_0}{\sigma},
\hspace*{0.25cm}\sigma=u_bc_0\\
\\
\DD{\phi}_0=\mu\DD{\Phi}_0\\
\\
(\nu R)^{-1}\DDD{\phi}_0= \rho R^{-1}\DDD{\Phi}_0
\\
\end{array}\right] & & \mbox{on}\hspace*{0.25cm} \eta=0\\
\\
\\
0=(\nu R)^{-1}\Phi_0^{iv} & & \mbox{in lower fluid}\\
\\
\phi_0=\D{\phi}_0=0, & & \mbox{on}\hspace*{0.25cm} \eta=-d^{-1}\\
\end{array}\right\}\label{2.17}
\end{eqnarray}
In the case of pPf, the system (\ref{2.17}) has a eigensolution
$$\Phi_0=(\eta-1)^2\left(\eta\df{d}{2\mu}\df{\mu-d^2}{1+d}+1\right)$$ and
$$\phi_0=(\eta+d^{-1})^2\left(\eta\df{d}{2}\df{\mu-d^2}{1+d}+d^{2}\right).$$

The eigenvalue $c_0$ is given by
\begin{eqnarray}
c_0=u_b(0)+\df{2\mu d(1+d)(\D{U}_b-\D{u}_b)}{4\mu d(1+d)^{2}+
(\mu-d^2)^2}\hspace*{0.5cm}\mbox{provided $\mu\ne 1$}\label{2.18}
\end{eqnarray}

Similarly, in the limit of slow basic flow, when the Reynolds number for the
basic flow is small, we may expand $c\sim R^{-1}c_0+c_1+...$, $\phi\sim\phi_0+R\phi_1+...$
and $\Phi\sim\Phi_0+R\Phi_1+...$ as $R\rightarrow 0$ for fixed ${\mathscr F}$ and ${\mathscr T}$. Thus
we obtain a system similar to (\ref{2.17}) whose solution is
\be
& &\Phi_0=A[\cosh k(\eta-1)-\cosh\lambda (\eta-1)]+B[\sinh \lambda(\eta-1)-\lambda k^{-1}\sinh k(\eta-1)]\no\\
& &\phi_0=a[\cosh k(\eta+d^{-1})-\cosh\gamma (\eta+d^{-1})]+b[\sinh\gamma(\eta+d^{-1})-\gamma k^{-1}\sinh k(\eta+d^{-1})]\no
\en
where $\lambda^2=k(k-ic_0)$ and $\gamma^2=k(k-i\nu c_0)$, and the eigenvalues $c_0$ and the constants $A,B, a$ and $b$ are
determined by applying the interface conditions at $\eta=0$. This yields a homogeneous $4\times 4$ set of equations whose
non-trivial solutions can only be obtained for certain values of $c_0$.

We next consider $O(k)$. To this order the system (\ref{2.15})
reduce to
\begin{eqnarray}
\left.\begin{array}{lll}
\Phi^{(1)}=\D{\Phi^{(1)}}=0, & & \mbox{on}\hspace*{0.25cm} \eta=1\\
\\
i(U_b-c_0)\DD{\Phi}_0-i\DD{U}_b\Phi_{0}=
R^{-1}\Phi^{iv}_1 & & \mbox{in upper fluid}\\
 & & \\
\left.\begin{array}{l}
\phi_0=\Phi_0\\
\\
\D{\phi}_1-\df{\D{u}_b}{\sigma}\left(\phi_{1}+\df{c_{1}}
{\sigma}\phi_{0}\right)=\\
\\
\hspace*{2.75cm}\D{\Phi}_1-\df{\D{U}_b}{\sigma}\left(\Phi_{1}+
\df{c_{1}}{\sigma}\Phi_{0}\right)\\
\\
\DD{\phi}_1=\mu\DD{\Phi}_1\\
\\
(\nu R)^{-1}\DDD{\phi}_1-i\left(\D{\phi}_1\sigma-\phi_{0}
\D{u}_b\right)+\df{i\phi_0}{\sigma}\df{{\sf F}}{R^2}=\\
\\
\hspace*{1cm}\rho R^{-1}\DDD{\Phi}_1-
\rho i\left(\D{\Phi}_1\sigma-\Phi_{0}\D{U}_b\right)+\\
\\
\hspace*{1cm}\df{\rho i \Phi_{0}{\sf F}}{\sigma R^2}\\
\\
\end{array}\right] & & \mbox{on}\hspace*{0.25cm} \eta=0\\
\\
i(u_b-c_{0})\DD{\phi}_0-i\DD{u}_b\phi_{0}=
(\nu R)^{-1}\phi^{iv}_1 & & \mbox{in lower fluid}\\
\\
\phi_{1}=\D{\phi}_1=0, & & \mbox{on}\hspace*{0.25cm} \eta=-d^{-1}\\
\end{array}\right\}\label{2.19}
\end{eqnarray}
Since $\phi^{(0)}, \Phi^{(0)}$ and $c^{(0)}$ are real, we can
easily see that $c^{(1)}$ is purely imaginary. Note that the
homogeneous part (\ref{2.19}) is the same as (\ref{2.17}). Thus,
in order that the inhomogeneous system to have a solution we must
satisfy certain solvability criterion. We introduce the adjoint
functions [6], $\Psi$ and $\psi$, and write down the system which
is adjoint to (\ref{2.19})
\begin{eqnarray}
\left.\begin{array}{lll}
\Psi=\D{\Psi}=0, & & \mbox{on}\hspace*{0.25cm} \eta=1\\
\\
0=R^{-1}\Psi^{iv} & & \mbox{in upper fluid}\\
\\
\left.\begin{array}{l} \psi=\Psi,\hspace*{0.5cm}
\D{\psi}=\D{\Psi},\hspace*{0.5cm}
\DD{\psi}=\mu\DD{\Psi}\\
\\
\DDD{\psi}-\df{\D{u}_b}{\sigma}\D{\psi}=\mu\left(
\DDD{\Psi}-\df{\D{U}_b}{\sigma}\D{\Psi}\right)\\
\\
\end{array}\right] & & \mbox{on}\hspace*{0.25cm} \eta=0\\
\\
0=(\nu R)^{-1}\psi^{iv} & & \mbox{in lower fluid}\\
\\
\psi=\D{\psi}=0, & & \mbox{on}\hspace*{0.25cm} \eta=-d^{-1}\\
\end{array}\right\}\label{2.20}
\end{eqnarray}
with the adjoint condition \footnote{Stuart's solvability
criterion states that if $L$ is a differential operator and $N$ is
its adjoint operator, then if $L\Phi=f$ and its adjoint is given
by $N\Psi=0$ the solvability condition states\\ $\int_{a}^{b}\Psi
L\Phi\,d\eta=\int_{a}^{b}\Phi\ub{N\Psi}{=0}\,d\eta$ which implies
$\int_{a}^{b}\Psi f\,d\eta=0.$}
\begin{eqnarray}
(\nu R)^{-1}\Int{-d^{-1}}{0}\psi\phi^{iv}_0\,d\eta+\rho R^{-1}
\Int{0}{1}\Psi\Phi^{iv}_0\,d\eta=(\nu
R)^{-1}\Int{-d^{-1}}{0}\phi_{0}\psi^{iv}\,d\eta +\rho R^{-1}
\Int{0}{1}\Phi_{0}\Psi^{iv}\,d\eta=0\no\\\no
\end{eqnarray}

The eigenvalue $c_{0}$ for the adjoint system (\ref{2.20}) is the
same as that of (\ref{2.17}), while the adjoint functions for pPf are
\begin{eqnarray}
& &\Psi=(\eta-1)^2\left[\eta\df{\mu-d^2+4d(1+d)}{2(\mu+d)}+1\right]\nonumber\\
\nonumber\\
& &\psi=(\eta+d^{-1})^2\left[\eta\df{d^2(\mu-d^2)-4\mu
d^2(1+d)}{2(\mu+d)}+d^2 \right]\nonumber
\end{eqnarray}
When the solvability condition is applied to (\ref{2.17}) we
obtain the following expression for $c_{1}$:
\begin{eqnarray}
\df{c_{1}\phi_{0}}{\sigma^2}(\nu R)^{-1}\DD{\psi}(\D{u}_b-
\D{U}_b)&=&-i\psi\left[(1-\rho)(\D{\phi}_0\sigma-\phi_{0}\D{u}_b)-
\df{\phi_{0}{\sf F}(1-\rho)}{\sigma R^2}\right]\nonumber\\
&+&i\Int{-d^{-1}}{0}\psi\left[(u_b-c_{0})\DD{\phi}_0-\DD{u}_b
\phi_{0}\right]\,d\eta\nonumber\\
&+&\rho i\Int{0}{1}\Psi\left[(U_b-c_{0})\DD{\Phi}_0-\DD{U}_b
\Phi_{0}\right]\,d\eta\label{2.22}
\end{eqnarray}
To $O(k)$, the condition for neutral stability is that $c_{1}=0$.
Hence, from (\ref{2.22}) we obtain the following relationship
between $k$ and $R$:
\begin{eqnarray}
\df{{\sf F}}{R^2}=\sigma(\D{\phi}_0\sigma-\D{u}_b)
&-&\sigma\left\{\ub{\df{1}{1-\rho}\Int{-d^{-1}}{0}\psi\left[(u_b-c_{0})
\DD{\phi}_0-\DD{u}_b\phi_{0}\right]\,d\eta}{I}\right.\nonumber\\
&+&\left.\ub{\rho\Int{0}{1}\Psi\left[(U_b-c_{0})\DD{\Phi}_0-\DD{U}_b
\Phi_{0}\right]\,d\eta}{J}\right\}\label{2.23}
\end{eqnarray}

Equation (\ref{2.23}) is a quadratic in $\sigma$ which may be
expressed as
\begin{eqnarray}
\D{\phi}_0\sigma^2-(\D{u}_b+I+J)\sigma-\df{{\sf F}}{R^2}=0
\nonumber
\end{eqnarray}
with its solutions given by
\begin{eqnarray}
\sigma=\df{(\D{u}_b+I+J)\pm\sqrt{(\D{u}_b+I+J)^2+ 4\D{\phi}_0{\sf
F}/R^2}}{2\D{\phi}_0}\equiv\gamma(R)\nonumber
\end{eqnarray}
Since $\sigma=u_b(0)-c$, we obtain the following compact relation
between the wavenumber $k$ and the Reynolds number $R$, i.e.
\begin{eqnarray}
k=\df{ng}{[u_b(0)-\gamma(R)]^2}\label{2.24}
\end{eqnarray}

\section{The short-wave limit}
We shall now focus our attention to conditions under which
short-wavelengths disturbances may grow along the interface. We
shall pose the following restrictions on the problem
\begin{eqnarray}
\left. \begin{array}{lclccclcl}
{\rm (a)} & & d\ll 1,   & & & & {\rm (c)} & & kd^3\gg 1\\
\\
{\rm (b)} & & k\ll 1,   & & & & {\rm (d)} & & kd\ll 1\\
\end{array} \right\} \label{4.0}
\end{eqnarray}

To study this, we will adopt an asymptotic analysis for pPf.
However,  before we commence with our analysis, it is convenient
to express the interfacial boundary conditions (\ref{2.15}) in
more compact mathematical form.  We do this by defining \be
\left.\begin{array}{l}
\Theta=U_b+\df{i\rho
k}{\sigma}\df{\sf F}{R^2}\\
\\
\Xi=\nu u_b+\df{1}{\sigma}({\sf F}+k^2{\sf
S})\left(\df{\nu}{R^2}+1\right)\\
\end{array}\right\}\label{4.1}
\en Thus, the second to the fourth interfacial conditions
(\ref{2.15}) may be written as \be \left.\begin{array}{l}
\Phi'-\phi'=\Phi(\Theta-\Xi)/\sigma\\
\\
\Phi''+k^2\Phi=\mu(\phi''+k^2\phi)\\
\\
-ikR(\sigma\Phi'+\Xi\Phi)-\Phi'''+3k^2\Phi'\\
\\
\hspace*{1cm}ikR\rho(\sigma\phi'+\Xi\phi)+\mu(\phi'''-3k^2\phi')\\
\\
\hspace*{1cm}=ikR({\sf F}+k^2{\sf S})\Phi/\sigma
\end{array}\right\}\mbox{on $\eta=0$}
 \label{4.2}
\en We shall also scale the coordinates $(\chi,\eta)$ with the
wavenumber $k$ such that \be
\theta=k\chi,\hspace*{1cm}\varphi=k\eta\label{4.3a} \en and assume
the following asymptotic expansions for $c, \phi$ and $\Phi$:
\begin{eqnarray}
\left.\begin{array}{l} c\sim
c_{0}+k^{-1}c_{1}+k^{-2}c_{2}+...\\
\\
\phi\sim\phi_{0}+k^{-1}\phi_{1}+k^{-2}\phi_{2}+...\\
\\
\Phi\sim\Phi_{0}+k^{-1}\Phi_{1}+k^{-2}\Phi_{2}+...\\
\end{array}\right\}\label{4.3b}
\end{eqnarray}

Neglecting terms of $O(d^{-1})$ by invoking the restriction (a) in
(24), replacing the independent variable by
$\hat{\varphi}=-\varphi$ and defining ${\mathscr
Y}=\hat{\varphi}/d$\,\,$(0\leq{\mathscr Y}\leq 1)$, the eigenfunction
for the lower fluid will satisfy the following Orr-Sommerfeld
equation \be \phi^{iv}-2(kd)^2\phi''+(kd)^4\phi=\df{iR\rho
d^3}{\mu^2}k\left\{\left[{\mathscr Y}-{\mathscr Y}^2+(1-c)\mu
d^{-1}\right]\left[\phi''-(kd)^2\phi\right]+2\phi\right\}\no\\
\label{4.5} \en in which we anticipate $c\sim1-O(k^{1/3})$.

Since the effective Reynolds number, $R\rho d^3/\mu^2$, is large,
we may neglect the viscous terms in (\ref{4.5}) and thus reduce it
to the Rayleigh equation \be {\mathscr Y}(1-{\mathscr
Y})\phi''+2\phi=0\label{4.6} \en Equation (\ref{4.6}) has a
regular solution \be \phi_2^o=K{\mathscr Y}(1-{\mathscr Y})\label{4.8} \en
and a solution with logarithmic branch point at ${\mathscr Y}=0$ and
${\mathscr Y}=1$, namely \be \phi_1^o=\Lambda\left\{{\mathscr Y}(1-{\mathscr
Y})\log{\mathscr Y}- \log(1-{\mathscr Y})+{\mathscr
Y}^2-\tf{1}{2}\right\}\label{4.9} \en where the superscript $o$ in
(\ref{4.8}) and (\ref{4.9}) denotes the outer solution.

Now to recover the boundary layer near the interface, and
following the usual practice in the theory of hydrodynamic
stability, we introduce $\eps=(iR\rho kd^3/\mu^2)^{-1/3}$ and let
${\sf Y}={\mathscr Y}/\eps$. Thus, (\ref{4.5}) becomes \be
\phi^{iv}-2(\eps kd)^2\phi''+(\eps kd)^4\phi= \left\{\left[{\sf
Y}(1-\eps{\sf Y})+(1-c)\mu/ \eps d\right] \left[\phi''-(\eps
kd)^2\phi\right]+2\eps\phi\right\}\hspace*{0.4cm} \label{4.10} \en
Expanding $\phi$ in powers of $\eps$ we obtain, at zero order in
$\eps$ \be \phi_0^{iv}-{\sf Y}\phi_0''=0\label{4.11} \en and at
$O(\eps)$ we get \be \phi_1^{iv}-{\sf Y}\phi_1''=-{\sf
Y}^2\phi_0''+2\phi_0\label{4.12} \en

The solution of (\ref{4.11}) is \be \phi^{I}=m_1+m_2{\sf
Y}+m_3{\mathscr A}_1({\sf Y},2)+m_4{\mathscr A}_2({\sf Y},2) \label{4.13}
\en In (\ref{4.13}), ${\mathscr A}_k({\sf Y},2),\,\,k=1,2$, denote
generalized Airy function \be {\mathscr A}_k({\sf Y},2)=\df{1}{2\pi
i}\int_{{\mathscr L}_k}t^{-2}\exp\left({\sf Y}t-
\tf{1}{3}t^3\right)\,dt\label{airy1} \en where ${\mathscr L}_k$ is
the appropriate path of integrals shown in figure 1.
\begin{figure}
\vspace{1pc}
    \begin{center}
        \includegraphics[width=6cm]{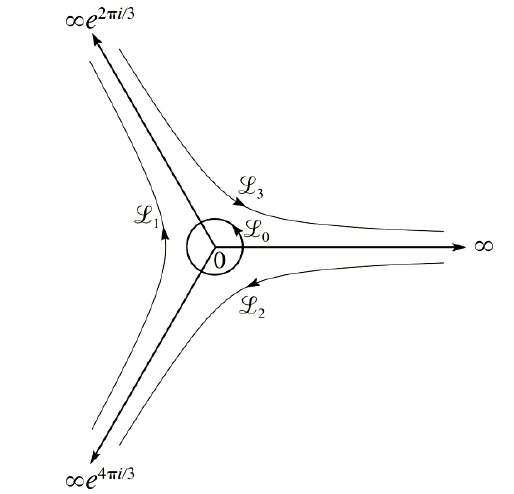}
    \end{center}
    \caption[oblique wave]{\footnotesize
The path ${\mathscr L}_0$, ${\mathscr L}_1$, ${\mathscr L}_2$ and
${\mathscr L}_3$ of integral (\ref{airy1}).} \label{fig:Fig1}
\end{figure}

For ${\sf Y}\gg 0$ the generalized Airy function ${\mathscr A}_2({\sf Y},2)$ is
exponentially large. Thus we set the constant $m_4=0$ in
(\ref{4.13}). The constants $m_1$ and $m_2$ are found by matching
(\ref{4.13}) with the inviscid solutions (\ref{4.8}) and
(\ref{4.9}). However, to do this we require the next-order
solutions to the constant and linear parts in (\ref{4.13}).

From (\ref{4.13}) we see that \be
& &\phi_1^I=m_1\left\{1+2\eps[n_1+n_2{\sf Y}+B_3({\sf Y},2,1)]\right\}\label{4.14}\\
\nonumber\\
& &\phi_2^I=m_2\sf{Y}(1-{\sf Y})\label{4.15} \en where the
superscript $I$ represent the inner solution and ${\mathscr B}_3({\sf
Y},2,1)$ is a generalized Airy function given by \be {\mathscr
B}_{3}({\sf Y},2,1)=\df{1}{2\pi i}\int_{\infty\exp[4\pi
i/3]}^{(0+)}t^{-2}\ln t \exp\left({\sf
Y}t-\tf{1}{3}t^3\right)\,dt\label{airy} \en

Now matching $\phi_2^I$ with $\phi_2^o$, and $\phi_1^I$ with
$\phi_1^o$, we obtain, respectively \be
m_1=-\Lambda/2,\,\,\,\,m_2=K\eps,
\,\,\,\,n_1=0,\,\,\,
n_2=-\log\eps-[\psi(2)+2\pi i]\label{4.18} \en where $\psi(2)$
is the psi function.

To evaluate the above integral we note that \be {\mathscr A}_1({\sf
Y},2)=-\df{1}{2}\left[{\sf Y}^2\int_{\infty}^{\sf
Y}\Ai(t)\,dt-{\sf Y}\Ai'({\sf Y})-\Ai({\sf Y})\right]\no \en where
$\Ai$ is the usual Airy function. Using the asymptotic
representation of the generalized Airy function
$${\mathscr A}_1(z,p)\sim\df{1}{2\sqrt{\pi}}z^{-(2p+3)/4}\exp\left(-\tf{2}{3}z^{3/2}\right)+...$$ and the following relationship
$${\mathscr A}_2(z,p)=e^{-2\pi i/3}{\mathscr A}_1\left(ze^{2\pi i/3},p\right)$$
it can be shown that
\be
\phi_1^{I}=-\eps\Lambda\left\{\df{1}{2\eps}-[\log\eps+\psi(2)+2\pi i]{\sf Y}-\df{{\sf Y}^{1/4}}{\sqrt{\pi}}\exp\left[-i\left(2{\sf Y}^{1/2}-\tf{3}{4}\pi
\right)\right]\right.& &\no\\
\left.-\df{{\sf Y}^{-5/4}}{\sqrt{\pi}}\exp\left(-\tf{2}{3}{\sf Y}^{3/2}\right)\right\}& &\no
\en
Note that, solutions of this kind are used as an initial guess for the full numerical integration.

\section{Nonlinear stability}
The governing equations and boundary conditions for the
two-dimensional motion of two superposed fluids in curvilinear
coordinates are
 \be \left.\begin{array}{c}
U=V=0\hspace*{0.3cm}{\rm on}\hspace*{0.3cm}\eta=1\\
\\
\PDD{U}{t}+J^{\half}\left\{U\PD{\chi}\left(J^{-\half}U\right)+V\PD{\eta}\left(J^{-\half}U\right)\right\}
+\tf{1}{2}J^{-1}\PDD{J}{\chi}(U^2+V^2)=\\
\\
\hspace*{1cm}-\PDD{P}{\chi}+R^{-1}\left\{\PDD{J}{\eta}\left[\PD{\eta}\left(J^{-\half}U\right)
-\PD{\chi}\left(J^{-\half}V\right)\right]+J\nabla^2\left(J^{-\half}U\right)\right\}\\
\\
\PDD{V}{t}+J^{\half}\left\{U\PD{\chi}\left(J^{-\half}V\right)+V\PD{\eta}\left(J^{-\half}V\right)\right\}
+\tf{1}{2}J^{-1}\PDD{J}{\eta}(U^2+V^2)=\\
\\
\hspace*{1cm}-\PDD{P}{\eta}+R^{-1}\left\{\PDD{J}{\chi}\left[\PD{\eta}\left(J^{-\half}U\right)
-\PD{\chi}\left(J^{-\half}V\right)\right]-J\nabla^2\left(J^{-\half}V\right)\right\}\\
\\
\PD{\chi}\left(J^{-\half}U\right)+\PD{\eta}\left(J^{-\half}V\right)=0\\
\\
\begin{array}{clc}
\left.\begin{array}{l} \hspace*{3cm}
u=U,\hspace*{0.25cm}v=V,\hspace*{0.25cm}
\PDD{\zeta}{t}+J^{\half}u\PDD{\zeta}{\chi}=v,\hspace*{0.25cm}\PDD{\zeta}{t}+J^{\half}U\PDD{\zeta}{\chi}=V\\
\\
\hspace*{1cm}
2J\zeta_\chi\PDD{v}{\eta}+\tf{1}{2}J^{\half}\left(1-J\zeta_\chi^2\right)\left(\PDD{u}{\eta}+\PDD{v}{\chi}\right)=
\mu\left[2J\zeta_\chi\PDD{V}{\eta}+\tf{1}{2}
\left(1-\zeta_\chi^2\right)\left(\PDD{U}{\eta}+\PDD{V}{\chi}\right)\right]\\
\\
\rho p-\df{\mathscr{F}}{R^2}\zeta-\df{2J^{\half}}{\nu
R\left(1-J\zeta_\chi^2\right)}\left[\left(1-J\zeta_\chi^2\right)\PDD{v}{\eta}-
J^{\half}\zeta_\chi\left(\PDD{u}{\eta}+\PDD{v}{\chi}\right)\right]+
\df{J\mathscr{T}\left(J^{-\half}J^{\half}_\chi\zeta_\chi+\zeta_{\chi\chi}\right)}{R^2\left(1+J\zeta_\chi^2\right)^{\thalf}}\\
\hspace*{1.5cm}=\rho\left\{P-\df{\mathscr{F}}{R^2}\zeta-\df{2J^{\half}}{\left(1+J\zeta_\chi^2\right)}
\left[\left(1-J\zeta_\chi^2\right)\PDD{V}{\eta}-
J^{\half}\zeta_\chi\left(\PDD{U}{\eta}+\PDD{V}{\chi}\right)\right]\right\}\\
\end{array}\right] & &\!\!\!\!
\mbox{on $\eta=0$}\end{array}\\
\\
\PDD{u}{t}+J^{\half}\left\{u\PD{\chi}\left(J^{-\half}u\right)+v\PD{\eta}\left(J^{-\half}v\right)\right\}
+\tf{1}{2}J^{-1}\PDD{J}{\chi}(u^2+v^2)=\\
\\
\hspace*{1cm}-\rho\PDD{p}{\chi}+(\nu
R)^{-1}\left\{\PDD{J}{\eta}\left[\PD{\eta}\left(J^{-\half}u\right)
-\PD{\chi}\left(J^{-\half}v\right)\right]+J\nabla^2\left(J^{-\half}u\right)\right\}\\
\\
\PDD{v}{t}+J^{\half}\left\{u\PD{\chi}\left(J^{-\half}v\right)+v\PD{\eta}\left(J^{-\half}v\right)\right\}
+\tf{1}{2}J^{-1}\PDD{J}{\eta}(u^2+v^2)=\\
\\
\hspace*{1cm}-\rho\PDD{p}{\eta}+(\nu
R)^{-1}\left\{\PDD{J}{\chi}\left[\PD{\eta}\left(J^{-\half}u\right)
-\PD{\chi}\left(J^{-\half}v\right)\right]-J\nabla^2\left(J^{-\half}v\right)\right\}\\
\\
\PD{\chi}\left(J^{-\half}u\right)+\PD{\eta}\left(J^{-\half}v\right)=0\\
\\u=v=0\hspace*{0.3cm}{\rm on}\hspace*{0.3cm}\eta=d^{-1}\\
\end{array}\right\}\,\,\,\,\,\,\,\label{nls1}
\en
where the upper and
lower case symbols refer to fluids above and below the interface,
respectively, and $\rho=\rho_a/\rho_w$; $\mu=\mu_a/\mu_w$;
$\nu=\nu_a/\nu_w$ and $d=D_a/D_w$, being the ratio of the height
above and below the interface. Here, $\rho$, $\mu$, and $\nu$ are
dimensionless density, dynamic viscosity and kinematic viscosity,
respectively, with understanding that the suffices $a$ and $w$
refer to air and water. Also, ${\mathscr F}=gD_a/U_\infty$,
${\mathscr T}=S/\rho_wU_\infty^2D_a$ and $R=U_\infty D_a/\nu_a$
are the Froude, Weber and Reynolds numbers respectively, with $g$
denoting the acceleration due to gravity, $S$ the coefficient of
surface tension, $U_\infty$ the free stream speed of the air, and
$\eta$ is the interface profile which is the Stokes wave.

For the disturbance near theocratical conditions we expand the
frequency and eigenfunction as a Taylor series. Thus
$$-ikc=-ik_cc-r+ia_1(k-k_c)-a_2(k-k_c)^2+d_1(R-R_c)+...$$
Now, since $d(kc)/dk|_{k_c}=-a_1$, then $-a_1$ is the group
velocity of the disturbance; $a_2$ is related to the curvature
of the nose of the neutral curve and $d_1$ is the exponential
growth of the wave for $R>R_c$. Likewise we expand $\Phi$ in the form
$$\Phi={\it\Phi}_1+(k-k_c){\it\Phi}_{10}+(R-R_c){\it\Phi}_{11}+(k-k_c)^2{\it\Phi}_{12}+...$$
with a similar expansion for $\phi$ in the lower fluid.

It is well known that the basic flow becomes unstable to
travelling wave disturbances when $R=R_c$ [3,8]. To examine the
growth of these Tollmien-Schlichting waves in the vicinity of
$R_c$ we introduce the small parameter \be
\eps=d_{1r}|R-R_c|\hspace*{0.3cm}{\rm
as}\hspace*{0.3cm}R\rightarrow R_c\label{nls2} \en and scaled
variable \be \tau=\eps
t,\hspace*{0.5cm}\xi=\eps^{\half}(\chi+a_1t)\label{nls3} \en

The continuity equations in (\ref{nls1}) are satisfied by the
streamfunctions $\Lambda$ and $\lambda$ for the upper and lower
fluids, respectively, thus \be
U=\PDD{\Lambda}{\eta},\,\,\,\,V=-\PDD{\Lambda}{\chi}
\hspace*{0.3cm}{\rm and}\hspace*{0.3cm}
u=\PDD{\lambda}{\eta},\,\,\,\,v=-\PDD{\lambda}{\chi}
\label{nls4} \en
Introducing the expansions
\be
& &\Lambda=\Phi_0(\eta,\xi,\tau)+[\Phi_1(\eta,\xi,\tau){\sf
E}+\Phi_2(\eta,\xi,\tau){\sf E}^2+{\rm c.c.}]+...\label{nls5u}\\
& &\lambda=\phi_0(\eta,\xi,\tau)+[\phi_1(\eta,\xi,\tau){\sf
E}+\phi_2(\eta,\xi,\tau){\sf E}^2+{\rm c.c.}]+...\label{nls5} \en
where c.c. denotes complex conjugate, and ${\sf
E}=\exp[ik_c(\chi-c_rt)].$ We further expand the functions in
(\ref{nls5}) as \be \phi_0(\eta,\xi,\tau)&=&\int_0^\infty
u_b\,d\eta+\eps\phi_{02}(\eta,\xi,\tau)+\eps^{\thalf}\phi_{03}(\eta,\xi,\tau)+\eps^2\phi_{04}(\eta,\xi,\tau)+
O\left(\eps^{\fhalf}\right)\no\\
\phi_1(\eta,\xi,\tau)&=&\eps^{\half}\phi_{11}(\eta,\xi,\tau)+\eps\phi_{12}(\eta,\xi,\tau)+
\eps^{\thalf}\phi_{13}(\eta,\xi,\tau)+\eps^{2}\phi_{14}(\eta,\xi,\tau)\no\\
&+&\eps^{\fhalf}\phi_{15}(\eta,\xi,\tau)+...\no
\no\\
\phi_2(\eta,\xi,\tau)&=&\eps\phi_{22}(\eta,\xi,\tau)+\eps^{\thalf}\phi_{23}(y\eta,\xi,\tau)+\eps^2\phi_{24}(\eta,\xi,\tau)+
O\left(\eps^{\fhalf}\right)\no\en

The pressure and interface displacement are expanded as \be
& &p=p_0(\chi,\eta,\xi,\tau)+[p_1(\eta,\xi,\tau){\sf E}+p_2(\eta,\xi,\tau){\sf E}^2+{\rm c.c.}]+...\label{nls10}\\
& &\zeta=\theta_0(\xi,\tau)+[\theta_1(\xi,\tau){\sf
E}+\theta_2(\xi,\tau)\sf{E}^2+{\rm c.c.}]+...\label{nls11} \en
where
\be p_0(\chi,\eta,\xi,\tau)&=&\df{u_b''\chi}{\mu
R_c}+\eps^{\half}p_{01}(\xi,\tau)+\eps\left\{p_{02}(\eta,\xi,\tau)-
\df{u_b''\chi}{\mu R_c^2d_{1r}}\right\}+...+O\left(\eps^{\thalf}\right)\no\\
\theta_0(\xi,\tau)&=&\eps\theta_{02}(\xi,\tau)+\eps^{\thalf}\theta_{03}(\xi,\tau)+\eps^2\theta_{04}(\xi,\tau)+
O\left(\eps^{\fhalf}\right)\no \en

Substituting in (\ref{nls1}) and equating powers of
$O(\eps^{\half}{\sf E})$ we obtain the coupled Orr-Sommerfeld
problem
\begin{eqnarray}
\left.\begin{array}{c}
{\it \Phi}_{11}=\D{{\it \Phi}}_{11}=0,\hspace*{0.3cm}{\rm on}\hspace*{0.3cm}\eta=1\\
\\
\mbox{in upper fluid}\\
\\
\mathscr{L}(k_c,{\it \Phi}_{11})=ik_c(U_b-c_r)(\DD{{\it
\Phi}}_{11}-k_c^2{\it \Phi}_{11})-ik_c\DD{U}_b{\it \Phi}_{11}-
R_c^{-1}({\it \Phi}_{11}^{iv}-2k_c^2\DD{\it \Phi}_{11}+k_c^4{\it
\Phi}_{11})=0\\
\\
\mbox{on $\eta=0$}\\
\\
\begin{array}{l}
\phi_{11}={\it\Phi}_{11}\\
\\
\D{\phi}_{11}-\df{\D{u}_b\phi_{11}}{\sigma_r}=
\D{{\it\Phi}}_{11}-\df{\D{U}_b{\it\Phi}_{11}}{\sigma_r}\\
\\
\DD{\phi}_{11}+k_c^2\phi_{11}=\mu(\DD{{\it\Phi}_{11}}+k_c^2{\it\Phi}_{11})\\
\\
{\sf q}(k_c,\phi_{11})=\df{1}{\nu
R_c}(\DDD{\phi}_{11}-3k_c^2\D{\phi}_{11})+ik_c(\phi_{11} u'_b-
\D{\phi}_{11}\sigma_r)+
i\df{k_c\phi_{11}}{\sigma_r}\df{(\mathscr{F}+k_c^2\mathscr{T})}{R_c^2}=\\
\hspace*{1cm}\df{\rho}{R_c}(\DDD{\it\Phi}_{11}-3k_c^2\D{\it\Phi}_{11})+
\rho ik_c({\it\Phi}_{11} U_b'-\D{\it\Phi}_{11}\sigma_r)+
\rho i\df{k_c{\it\Phi}_{11}}{\sigma_r}\df{{\mathscr{F}}}{R_c^2}=\rho\mathscr{Q}(k_c,{\it\Phi}_{11})\\
\end{array}\\
\\
\mbox{in lower fluid}\\
\\
{\frak l}(k_c,{\it
\phi}_{11})=ik(u_b-c_r)(\DD{\phi}_{11}-k_c^2\phi_{11})-ik_c\DD{u}_b\phi_{11}-(\nu
R_c)^{-1}(\phi_{11}^{iv}-2k_c^2\DD{\phi}_{11}+k_c^4\phi_{11})=0 \\
\\
\phi_{11}=\D{\phi}_{11}=0,\hspace*{0.3cm}{\rm on}\hspace*{0.3cm}\eta=-d^{-1}\\
\end{array}\right\}\no\\
\label{nls14}
\end{eqnarray}
Note that the system (\ref{nls14}) is the same as (\ref{2.15}) with
$R=R_c$, $k=k_c$ and $c=c_r$. The solution of (\ref{nls14}) may be written as \be
\left.\begin{array}{l}
{\it\Phi}_{11}=A(\tau,\xi){\it\Psi}_1(\eta)\\
\\
{\it\phi}_{11}=A(\tau,\xi){\it\psi}_1(\eta)\end{array}\right\} \label{nls15} \en
where $A(\tau,\xi)$ is the amplitude function to be determined.

The distortion of the mean flow occurs at $O(\eps{\sf E}^0)$,
which is
\begin{eqnarray}
\left.\begin{array}{c}

{\it \Phi}_{02}=\D{{\it \Phi}}_{02}=0,\hspace*{0.3cm}{\rm on}\hspace*{0.3cm}\eta=1\\
\\
R_c^{-1}{\it \Phi}_{02}^{iv}=ik_c|A|^2(\DD{\it
\Psi}_{1}\ts{\it\Psi}_1-{\it \Psi}_{1}\ts{\it\Psi}'_1)''\hspace*{0.3cm}\mbox{in upper fluid}\\
\\
\mbox{on $\eta=0$}\\
\\
\begin{array}{l}
{\phi}_{02}+\psi_1'\ts{\zeta}_1|A|^2={\it\Phi}_{02}+{\it\Psi}_1'\ts{\zeta}_1|A|^2\\
\\
\phi_{02}''-\df{u_b'\phi_{02}}{u_b+a_1}+|A|^2\left\{u_b''|\zeta_1|^2-\df{u_b'}{u_b+a_1}
(\ts{u}_{11}\zeta_1+\psi_1'\ts{\zeta}_1)+[\zeta_1\ts{\psi}_1+{\rm c.c.}]\right\}=\\
\\
\hspace*{0.5cm}{\it\Phi}_{02}''-\df{U_b'{\it\Phi}_{02}}{U_b+a_1}+|A|^2\left\{U_b''|\zeta_1|^2-\df{U_b'}{U_b+a_1}
(\ts{U}_{11}\zeta_1+{\it\Psi}_1'\ts{\zeta}_1)+[\zeta_1\ts{\it\Psi}_1+{\rm c.c.}]\right\}\\
\\
\phi_{02}''-\nu
R_cik_c|A|^2(\ts{\psi}_1\psi_1'-\psi_1\ts{\psi}_1')=
\mu\left[{\it\Phi}_{02}''-R_cik_c|A|^2(\ts{\it\Psi}_1{\it\Psi}_1'-{\it\Psi}_1\ts{\it\Psi}_1')\right]\\
\\
(\nu
R_c)^{-1}\DDD{\it\phi}_{02}-ik_c|A|^2(\ts{\psi}_1\psi_1'-\psi_1\ts{\psi}_1')'=
\rho\left[R_c^{-1}\DDD{\it\Phi}_{02}-ik_c|A|^2(\ts{\it\Psi}_1{\it\Psi}_1'-{\it\Psi}_1\ts{\it\Psi}_1')'\right]\\
\end{array} \\
\\
(\nu R_c)^{-1}{\it \phi}_{02}^{iv}=ik_c|A|^2(\DD{\it
\psi}_{1}\ts{\it\psi}_1-{\it
\psi}_{1}\ts{\it\psi}'_1)''\hspace*{0.3cm}\mbox{in lower fluid}
\\
\\
{\it \phi}_{02}=\D{{\it \phi}}_{02}=0,\hspace*{0.3cm}{\rm on}\hspace*{0.3cm}\eta=-d^{-1}\\
\end{array}\right\}\no\\
\label{nls14a}
\end{eqnarray}
where \be
\left.\begin{array}{l}
\zeta_1=-\psi_1(0)/\sigma_r=-{\it\Psi}_1(0)/\sigma_r\\
\\
u_{11}=\psi_1'(0)+\zeta_1u_b'(0)\\
\\
U_{11}={\it\Psi}_1'(0)+\zeta_1U_b'(0)\\
\\
ik_c\rho\B{p}_{11}'=(k_c^2/\nu
R_c)[\psi_1''(0)-k_c^2\psi_1(0)]-ik_c^3\sigma_r\psi_1(0)\\
\\
ik_c\B{P}_{11}'=(k_c^2/R_c)[{\it\Psi}_1''(0)-k_c^2{\it\Psi}_1(0)]-ik_c^3\sigma_r{\it\Psi}_1(0)\\
\end{array}\right\}\label{nls15}
\en and a tilde denotes complex conjugate. The system
(\ref{nls14}) has unique solution of the form \be
\left.\begin{array}{l}
{\it\Phi}_{02}={\frak F}(\eta)|A|^2\\
\\
{\it\phi}_{02}={\frak f}(\eta)|A|^2\\
\end{array}\right\}\label{nls16}
\en
where
\be
\begin{array}{lcl}
{\frak
f}(\eta)=\Int{-d^{-1}}{\eta}s(y)\,dy+(\eta+d^{-1})^2(b\eta+c), & &
{\frak
F}(\eta)=\Int{1}{y}S(y)\,dy+(\eta-1)^2(B\eta+C)\\
\\
s(\eta)=ik_c\nu R_c\Int{-d^{-1}}{\eta}(\ts{\psi}_1\psi_1'-\psi_1\ts{\psi}_1')\,dy, & &
S(\eta)=ik_cR_c\Int{1}{\eta}(\ts{\it\Psi}_1{\it\Psi}_1'-{\it\Psi}_1\ts{\it\Psi}_1')\,dy\\
\end{array}\no
\en The constants $B,C,b$ and $c$ are found by applying the
interface conditions.

At $O(\eps{\sf{E}}^2)$ we obtain the system governing the second
harmonic, namely
\begin{eqnarray}
\left.\begin{array}{c}
{\it \Phi}_{22}=\D{{\it \Phi}}_{22}=0,\hspace*{0.3cm}{\rm on}\hspace*{0.3cm}\eta=1 \\
\\
{\mathscr{L}}(2k_c,{\it\Phi}_{22})=ik_cA^2({\it
\Psi}_{1}{\it\Psi}_1'''-{\it \Psi}_{1}'{\it\Psi}'_1)\hspace*{0.3cm}\mbox{in upper fluid}\\
\\
\mbox{on $\eta=0$}\\
\\
\begin{array}{l}
{\phi}_{22}+\tf{1}{2}{\zeta}_1\psi_1'A^2={\it\Phi}_{22}+\tf{1}{2}{\zeta}_1{\it\Psi}_1'A^2\\
\\
\phi_{22}'-\df{u_b'\phi_{22}}{\sigma_r}+A^2\left[\zeta_1\psi_1''-\tf{1}{2}u_b'\zeta_1(u_{11}+\psi_1')+\tf{1}{2}u_b''\zeta_1^2\right]=\\
\\
\hspace*{0.5cm}{\it\Phi}_{22}'-\df{U_b'{\it\Phi}_{22}}{\sigma_r}+A^2\left[\zeta_1{\it\Psi}_1''-
\tf{1}{2}U_b'\zeta_1(U_{11}+{\it\Psi}_1')+\tf{1}{2}U_b''\zeta_1^2\right]\\
\\
\phi_{22}''+4k_c^2\phi_{22}+A^2\zeta_1\left[\psi_1'(8k_c^2+\nu
R_cik_c\sigma_r)-
\nu R_cik\psi_1\left(u_b'+\df{\mathscr{F}+k_c^2\mathscr{T}}{\sigma_rR_c^2}\right)\right]=\\
\hspace*{0.6cm}\mu\left\{{\it\Phi}_{22}''+4k_c^2{\it\Phi}_{22}+A^2\zeta_1\left[{\it\Psi}_1'(8k_c^2+R_cik_c\sigma_r)-
R_cik{\it\Psi}_1\left(U_b'+\df{\mathscr{F}}{\sigma_rR_c^2}\right)\right]\right\}\\
\\
{\sf
q}(2k_c,\phi_{22})+A^2\left[ik_c(\psi_1\psi_1''-\psi_1'\psi_1')+
\df{ik_c\zeta_1(u_{11}+\psi_1')(\mathscr{F}+4k_c^2\mathscr{T})}{\sigma_rR_c^2}\right.\\
\hspace*{0.6cm}\left.+\zeta_1\left(2ik_c\rho\B{p}'_{11}-\df{4k_c^2\psi_1''}{\nu R_c}\right)\right]\\
\\
\hspace*{0.6cm}=\rho\left\{{\mathscr{Q}}(2k_c,{\it\Phi}_{22})+A^2\left[ik_c({\it\Psi}_1{\it\Psi}_1''-
{\it\Psi}_1'{\it\Psi}_1')+\df{ik_c\zeta_1(U_{11}+{\it\Psi}_1')\mathscr{F}}
{\sigma_rR_c^2}\right.\right.\\
\hspace*{0.6cm}\left.\left.+\zeta_1\left(2ik_c\B{P}'_{11}-\df{4k_c^2\psi_1''}{R_c}\right)\right]\right\}
\\
\end{array}\\
\\
{\frak l}(2k_c,{\it\phi}_{22})=ik_cA^2({\it
\psi}_{1}{\it\psi}_1'''-{\it
\psi}_{1}'{\it\psi}'_1)\hspace*{0.3cm}\mbox{in lower fluid}\\
\\
{\it \phi}_{22}=\D{{\it \phi}}_{22}=0,\hspace*{0.3cm}\mbox{on $\eta=-d^{-1}$} \\
\end{array}\right\}\no\\
\label{nls17}
\end{eqnarray}
The solution of (\ref{nls17}) may be expressed as \be
\left.\begin{array}{l}
{\it\Phi}_{22}=A^2{\it\Psi}_2\\
\\
{\it\phi}_{22}=A^2{\it\psi}_2\\
\end{array}\right\}
\label{nls18} \en with the assumption
that the solution is unique.

We emphasize that the adjoint equation in the upper and he lower
fluid layer is the adjoint of the Orr-Sommerfeld equation for a
homogeneous fluid. The adjoint system possesses the following
relationship
$$\int_{-d^{-1}}^0\psi{\frak l}(\phi)\,d\eta+\rho\int_0^1\Psi{\mathscr L}(\Phi)\,d\eta=
\int_{-d^{-1}}^0\phi{\frak
l}^+(\psi)\,d\eta+\rho\int_0^1\Phi{\mathscr L}^+(\Psi)\,d\eta$$
where the operators ${\mathscr L}$ and ${\frak l}$ are the same as
that given in (\ref{nls14}), and \be
& &{\mathscr L}^+(\Psi)=ik_c(U_b-c_r)(\Psi''-k_c^2\Psi)+2ik_cU_b'\Psi'-R_c^{-1}(\Psi^{iv}-2k_c^2\Psi''+k_c^4\Psi)\no\\
\no\\
& &{\frak l}^+(\psi)=ik_c(u_b-c_r)(\psi''-k_c^2\psi)+2ik_cu_b'\psi'-(\nu R_c)^{-1}(\psi^{iv}-2k_c^2\psi''+k_c^4\psi)\no
\en

The system of equations for $O(\eps{\sf E})$ is given by
\begin{eqnarray}
\left.\begin{array}{c}
{\it \Phi}_{12}=\D{{\it \Phi}}_{12}=0,\hspace*{0.3cm}\mbox{on
$\eta=1$} \\
\\
{\mathscr{L}}(k_c,{\it\Phi}_{12})=\PDD{A}{\xi}\left\{{\it
\Psi}_{1}[U_b''+2k_c^2(U_b-c_r)]-[(U_b+a_1)-4ik_c/R_c]({\it\Psi}_1''-k_c^2{\it
\Psi}_{1})\right\},\hspace*{0.3cm}\mbox{in
UF}\\
\\
\mbox{on $\eta=0$}\\
\\
\begin{array}{l}
{\phi}_{12}={\it\Phi}_{12}\\
\\
\phi_{12}'-\df{u_b'\phi_{12}}{\sigma_r}-i\PDD{A}{\xi}\df{u_b(a_1+c_r)}{k_c\sigma_r^2}\psi_1=
{\it\Phi}_{12}'-\df{U_b'{\it\Phi}_{12}}{\sigma_r}-i\PDD{A}{\xi}\df{U_b(a_1+c_r)}
{k_c\sigma_r^2}{\it\Psi}_1\\
\\
{\sf q}(k_c,\phi_{12})+\PDD{A}{\xi}\left\{\df{6ik_c}{\nu
R_c}-[\psi_1'(u_b-a_1)-\psi_1u_b']\right.\\
\hspace*{0.5cm}\left.-\psi_1
\df{\mathscr{F}(-u_b+2c_r+a_1)-k_c^2\mathscr{T}(3u_b-4c_r-a_1)}{R_c^2\sigma_r^2}\right\}\\
\hspace*{0.5cm}={\mathscr{Q}}(k_c,{\it\Phi}_{12})+\PDD{A}{\xi}\left\{\df{6ik_c}{\nu
R_c}-[{\it\Psi}_1'(U_b-a_1)-{\it\Psi}_1U_b']-{\it\Psi}_1
\df{\mathscr{F}(-U_b+2c_r+a_1)}{R_c^2\sigma_r^2}\right\}
\\
\end{array}\\
\\
{\frak l}(k_c,{\it\phi}_{12})=\PDD{A}{\xi}\left\{{\it
\psi}_{1}[u_b''+2k_c^2(u_b-c_r)]-[(u_b+a_1)-4ik_c/R_c]({\it\psi}_1''-k_c^2{\it
\psi}_{1})\right\},\hspace*{0.3cm}\mbox{in
LF}\\
\\
{\it \phi}_{12}=\D{{\it \phi}}_{12}=0,\hspace*{0.3cm}\mbox{on $\eta=-d^{-1}$}\\
\end{array}\right\}\no\\
\label{nls19}
\end{eqnarray}
where UF and LF stands for upper and lower fluid, respectively.
The solution of the $O(\eps{\sf E})$ system is \be
\left.\begin{array}{l}
{\it\Phi}_{12}=-i\PDD{A}{\xi}{\it\Psi}_{10}+A_2{\it\Psi}_{1}\\
\\
{\it\phi}_{12}=-i\PDD{A}{\xi}{\it\psi}_{10}+A_2{\it\psi}_{1}\\
\end{array}\right\}
\label{nls20} \en where $A_2$ is and arbitrary function of $\tau$
and $\xi$, and ${\it\Psi}_{10}$ and ${\it\psi}_{10}$ are the
\nopagebreak solutions of the following system
\begin{eqnarray}
\left.\begin{array}{c}
\\
{\it \Psi}_{10}=\D{{\it \Psi}}_{10}=0,\hspace*{0.3cm}\mbox{on
$\eta=1$}\\
\\
{\mathscr{L}^+}(k_c,{\it\Psi}_{10})=i{\it\Psi}_1[U_b''+2k_c^2(U_b-c_r)]-[4ik_c/R_c+i(U_b+a_1)]({\it\Psi}_1''-k_c^2{\it
\Psi}_{1}),\hspace*{0.3cm}\mbox{in
UF}\\
\\
\mbox{on $\eta=0$}\\
 \begin{array}{l}
{\psi}_{10}={\it\Psi}_{10}\\
\\
\psi_{10}'-\df{u_b'\psi_{10}}{\sigma_r}+\df{u_b(a_1+c_r)}{k_c\sigma_r^2}\psi_1=
{\it\Psi}_{10}'-\df{U_b'{\it\Psi}_{10}}{\sigma_r}+\df{U_b(a_1+c_r)}
{k_c\sigma_r^2}{\it\Psi}_1\\
\\
\psi_{10}''+k_c^2\psi_{10}+2k_c\psi_1={\it\Psi}_{10}''+k_c^2{\it\Psi}_{10}+2k_c{\it\Psi_1}\\
\\
{\sf q}(k_c,\phi_{10})-\df{6k_c}{\nu
R_c}-i[\psi_1'(u_b-a_1)-\psi_1u_b']-i\psi_1
\df{\mathscr{F}(-u_b+2c_r+a_1)-k_c^2\mathscr{T}(3u_b-4c_r-a_1)}{R_c^2\sigma_r^2}\\
\\
\hspace*{0.5cm}={\mathscr{Q}}(k_c,{\it\Psi}_{10})-\df{6ik_c}{
R_c}-i[{\it\Psi}_1'(U_b-a_1)-{\it\Psi}_1U_b']-i{\it\Psi}_1
\df{\mathscr{F}(-U_b+2c_r+a_1)}{R_c^2\sigma_r^2}
\\
\end{array}\\
\\
{\frak l}^+(k_c,{\it\psi}_{10})=i{\it
\psi}_{1}[u_b''+2k_c^2(u_b-c_r)]-\left[\df{4ik_c}{\nu
R_c}+i(u_b+a_1)\right]({\it\psi}_1''-k_c^2{\it
\psi}_{1}),\hspace*{0.3cm}\mbox{in
LF}\\
\\
{\it \psi}_{10}=\D{{\it \psi}}_{10}=0,\hspace*{0.3cm}\mbox{on $\eta=-d^{-1}$}\\
\end{array}\right\}\no\\
\label{nls21}
\end{eqnarray}
since the adjoint to the system (\ref{2.15}), at critical conditions, satisfies
\begin{eqnarray}
\left.\begin{array}{lll}
\Psi=\D{\Psi}=0, & & \mbox{on}\hspace*{0.25cm} \eta=1\\
\\
{\mathscr L}^+(\Psi)=0 & &
\mbox{in upper fluid}\\
\\
\left.\begin{array}{l}
\psi=\Psi\\
\\
\D{\psi}=\D{\Psi}\\
\\
\DD{\psi}+k_c^2\psi=\mu(\DD{\Psi}+k_c^2\Psi)\\
\\
-\df{\DDD{\psi}-3k_c^2\psi}{\nu R_c}+\df{u_b'(\psi''+k_c^2\psi)}{\sigma_r\nu R_c}+ik_c\sigma_r\psi'-\df{ik_c\psi({\sf F}+k_c^2{\sf T})}{\sigma_r R_c^2}=\\
\\
\rho\left[-\df{\DDD{\Psi}-3k_c^2\Psi}{R_c}+\df{U_b'(\Psi''+k_c^2\Psi)}{\sigma_r R_c}+ik_c\sigma_r\Psi'-\df{ik_c\Psi{\sf F}}{\sigma_r R_c^2}\right]\\
\end{array}\right] & & \mbox{on}\hspace*{0.25cm} \eta=0\\
\\
{\frak l}^+(\psi)=0 & &
\mbox{in lower fluid}\\
\\
\psi=\D{\psi}=0, & & \mbox{on}\hspace*{0.25cm} \eta=-d^{-1}\\
\end{array}\right\}\no\\\label{adjor}
\end{eqnarray}

Next, at $O\left(\eps^{\thalf}{\sf E}\right)$, we obtain the
system of equations
\begin{eqnarray}
\left.\begin{array}{c}
\\
{\it \Phi}_{13}=\D{{\it \Phi}}_{13}=0,\hspace*{0.3cm}\mbox{on
$\eta=1$}\\
\\
{\mathscr{L}}(k_c,{\it\Psi}_{13})=(\D{\it\Psi}_1-k_c^2{\it\Psi}_1)\left(a_2\PDDT{A}{\xi}-\PDD{A}{\tau}\right)-
\df{A({\it\Psi}_{1}^{iv}-2k_c^2\DD{\it\Psi}_1+k_c^4{\it\Psi}_1)}{d_{1r}R_c^2}\\
\hspace*{0.5cm}-{\mathscr
M}\left(i\PDD{A_2}{\xi}+\PDDT{A}{\xi}\right)+A|A|^2{\mathscr
G}(\eta),\hspace*{0.3cm}\mbox{in
UF}\\
\\
\mbox{on $\eta=0$}\\
\\
 \begin{array}{l}
\phi_{13}-{\sf w}A|A|^2={\it\Phi}_{13}-{\mathscr W}A|A|^2,\\
\\
\phi_{13}'-\df{u_b'\phi_{13}}{\sigma_r} + \left(a_2\df{\partial^2
A}{\partial \xi^2}- \df{\partial A}{\partial
\tau}\right)\df{iu_b'\psi_1}{k_c\sigma_r^2} - {\sf
m}\left(i\df{\partial A_2}{\partial \xi}
+ \df{\partial^2 A}{\partial \xi^2}\right)\\
\hspace*{0.5cm}+A|A|^2\left[{\mathscr U} +\df{u_b'}{\sigma_r}(K_l)\right]\\
\hspace*{0.5cm}={\it\Phi}_{13}'-\df{U_b'{\it\Phi}_{13}}{\sigma_r} +
\left(a_2\df{\partial^2 A}{\partial \xi^2}- \df{\partial
A}{\partial \tau}\right)\df{iU_b'{\it\Psi}_1}{k_c\sigma_r^2} -
{\mathscr M}\left(i\df{\partial A_2}{\partial \xi}
+ \df{\partial^2 A}{\partial \xi^2}\right)\\
\hspace*{0.5cm}+A|A|^2\left[{\mathscr U} +\df{U_b'}{\sigma_r}(K_u)\right]\\
\\
\phi_{13}''+k_c^2\phi_{13} - {\sf m}\left(i \df{\partial
A_2}{\partial
\xi}+\df{\partial^2 A}{\partial \xi^2}\right)+A|A|^2{\sf s}\\
\hspace*{0.5cm}=\mu\left[{\it\Phi}_{13}''+k_c^2{\it\Phi}_{13} - {\mathscr M}\left(i
\df{\partial A_2}{\partial \xi}+\df{\partial^2 A}
{\partial \xi^2}\right)+A|A|^2{\mathscr S}\right],\\
\\
{\sf q}(k_c,\phi_{13})+\left(a_2\df{\partial^2 A}{\partial \xi^2}-
\df{\partial A}{\partial \tau}\right)\left[
\psi_1'+\psi_1\df{(\mathscr{F}+k_c^2\mathscr{T})}{\sigma_r
R_c^2}\right] - {\sf m}\left(\df{\partial A_2}{\partial \xi}
+ i\df{\partial^2 A}{\partial \xi^2}\right)\\
\hspace*{0.5cm}+\df{Aik_c}{d_{1r}R_c}\left(
u_b'\psi_1-\sigma_r\psi_1'-\psi_{1}\df{\mathscr{F}+k_c^2\mathscr{T}}{\sigma_rR_c^2}
\right) + {\sf n}A|A|^2,\\
\hspace*{0.5cm}=\rho\left\{\left(a_2\df{\partial^2 A}{\partial \xi^2}-
\df{\partial A}{\partial \tau}\right)\left[
{\it\Psi}_1'+{\it\Psi}_1\df{\mathscr{F}}{\sigma_r R_c^2}\right] - {\mathscr
M}\left(\df{\partial A_2}{\partial \xi}
+ i\df{\partial^2 A}{\partial \xi^2}\right)\right.\\
\left.\hspace*{0.5cm}+\df{Aik_c}{d_{1r}R_c}\left(
U_b'{\it\Psi}_1-\sigma_r{\it\Psi}_1'-{\it\Psi}_{1}\df{\mathscr{F}}{\sigma_rR_c^2}
\right) + {\mathscr N}A|A|^2\right\},\\
\\
\end{array}\\
{\frak
l}(k_c,{\it\psi}_{13})=(\D{\psi}_1-k_c^2\psi_1)\left(a_2\PDDT{A}{\xi}-\PDD{A}{\tau}\right)-
\df{A(\psi_{1}^{iv}-2k_c^2\DD{\psi}_1+k_c^4\psi_1)}{d_{1r}\nu R_c^2}\\
\hspace*{0.5cm}-{\sf
m}\left(i\PDD{A_2}{\xi}+\PDDT{A}{\xi}\right)+A|A|^2{\sf
g}(\eta),\hspace*{0.3cm}\mbox{in
LF}\\
\\
{\it \psi}_{13}=\D{{\it \psi}}_{13}=0,\hspace*{0.3cm}\mbox{on $\eta=-d^{-1}$}\\
\end{array}\right\}\no\\
\label{nls22}
\end{eqnarray}
where
\be
& &{\mathscr U}=\zeta_{02}\psi_1''+{\frak f}''\zeta_1+\zeta_{22}\ts{\psi}_1''+\ts{\zeta}_1\psi_2'+\psi_1'''|\zeta_1|^2
+u_b''(\zeta_{02}\zeta_1+\zeta_{22}\ts{\zeta}_1)+\tf{1}{2}\ts{\psi}_1'''\zeta_1^2\no\\
& &{\mathscr U}=\zeta_{02}{\it\Psi}_1''+{\frak F}''\zeta_1+\zeta_{22}\ts{\it\Psi}_1''+\ts{\zeta}_1{\it\Psi}_2'+{\it\Psi}_1'''|\zeta_1|^2
+U_b''(\zeta_{02}\zeta_1+\zeta_{22}\ts{\zeta}_1)+\tf{1}{2}\ts{\it\Psi}_1'''\zeta_1^2\no
\en

Proceeding as above we obtain the following set of equations at
$O\left(\eps^{\fhalf}{\sf E}^2\right)$.
\begin{eqnarray}
\left.\begin{array}{c}
\\
{\it \Phi}_{23}=\D{{\it \Phi}}_{23}=0,\hspace*{0.3cm}\mbox{on
$\eta=1$}\\
\\
{\mathscr{L}}(k_c,{\it\Phi}_{23})=
\left(a_2\PDDT{A}{\xi}-\PDD{A}{\tau}\right)(\DD{\it\Psi}_1-k_c^2{\it\Psi}_1)-
\df{A({\it\Psi}_{1}^{iv}-2k_c^2\DD{\it\Psi}_1+k_c^4{\it\Psi}_1)}{d_{1r}R_c^2}\\
\hspace*{0.5cm} -i{\mathscr \alpha}_u\PDD{A_2}{\xi}-{\mathscr
\beta}_u\PDDT{A}{\xi} +A|A|^2{\mathscr G}(\eta) + A|A|^4{\mathscr
H}(\eta),\hspace*{0.3cm}\mbox{in
UF}\\
\\
\mbox{on $\eta=0$}\\
\\
 \begin{array}{l}
\phi_{23}-{\sf v}A|A|^2-{\sf w}A|A|^4={\it\Phi}_{23}-{\mathscr V}A|A|^2-{\mathscr W}A|A|^4,\\
\\
\phi_{23}'-\df{u_b'\phi_{23}}{\sigma_r}
\left(a_2\df{\partial^2 A}{\partial \xi^2}-\df{\partial A}{\partial \tau}\right)
\df{iu_b'\psi_1}{k_c\sigma_r^2} -
-i{\mathscr \alpha}_l\PDD{A_2}{\xi}-{\mathscr \beta}_l\PDDT{A}{\xi}\\
\hspace*{0.5cm}+A|A|^2\left[{\mathscr U} +\df{u_b'}{\sigma_r}(K_l)\right]
+A|A|^4\left[{\mathscr U} +\df{u_b'}{\sigma_r}(J_l)\right]\\
\hspace*{0.5cm}={\it\Phi}_{23}'-\df{U_b'{\it\Phi}_{23}}{\sigma_r}
\left(a_2\df{\partial^2 A}{\partial \xi^2}-
\df{\partial A}{\partial \tau}\right)
\df{iU_b'{\it\Psi}_1}{k_c\sigma_r^2}
-i{\mathscr \alpha}_u\PDD{A_2}{\xi}\\
\hspace*{0.5cm}-{\mathscr \beta}_u\PDDT{A}{\xi}
+A|A|^2\left[{\mathscr U} +\df{U_b'}{\sigma_r}(K_u)\right]
+A|A|^4\left[{\mathscr U} +\df{U_b'}{\sigma_r}(J_u)\right],\\
\\
\phi_{23}''+k_c^2\phi_{23} -i{\mathscr \alpha}_l \df{\partial A_2}{\partial \xi}
+A|A|^2{\sf s} + A|A|^4{\sf r}\\
\hspace*{0.5cm}=\mu\left[{\it\Phi}_{23}''+k_c^2{\it\Phi}_{23} - i{\mathscr \alpha}_u \df{\partial A_2}{\partial \xi}
+A|A|^2{\mathscr S} +A|A|^4{\mathscr R}\right],\\
\\
{\sf q}(k_c,\phi_{23})+\left(a_2\df{\partial^2 A}{\partial \xi^2}-
\df{\partial A}{\partial \tau}\right)\left[
\psi_1'+\psi_1\df{(\mathscr{F}+k_c^2\mathscr{T})}{\sigma_r^2
R_c^3}\right] - i{\mathscr \alpha}_l\df{\partial A_2}{\partial \xi}
- {\mathscr \beta}_l\df{\partial^2 A}{\partial \xi^2}\\
\hspace*{0.5cm}+\df{Aik_c}{d_{1r}R_c}\left(
u_b'\psi_1-\sigma_r\psi_1'-\psi_{1}\df{\mathscr{F}+k_c^2\mathscr{T}}{\sigma_rR_c^2}
\right) + {\sf n}A|A|^2 + {\sf m}A|A|^4\\
\hspace*{0.5cm}=\rho\left\{{\sf Q}(k_c,{\it\Phi}_{23})+\left(a_2\df{\partial^2 A}{\partial \xi^2}-
\df{\partial A}{\partial \tau}\right)\left[
{\it\Psi}_1'+{\it\Psi}_1\df{\mathscr{F}}{\sigma_r R_c^2}\right] -
i{\mathscr \alpha}_u\df{\partial A_2}{\partial \xi}
- {\mathscr \beta}_u\df{\partial^2 A}{\partial \xi^2}\right.\\
\left.\hspace*{0.5cm}+\df{Aik_c}{d_{1r}R_c}\left(
U_b'{\it\Psi}_1-\sigma_r{\it\Psi}_1'-{\it\Psi}_{1}\df{\mathscr{F}}{\sigma_rR_c^2}
\right) + {\mathscr N}A|A|^2 + {\mathscr M}A|A|^4\right\},\\
\\
\end{array}\\
{\frak l}(k_c,{\it\psi}_{23})=(\DD{\psi}_1-k_c^2\psi_1)\left(a_2\PDDT{A}{\xi}-\PDD{A}{\tau}\right)-
\df{A(\psi_{1}^{iv}-2k_c^2\DD{\psi}_1+k_c^4\psi_1)}{d_{1r}\nu R_c^2}\\
\hspace*{0.5cm}-i{\mathscr \alpha}_l\PDD{A_2}{\xi}-{\mathscr
\beta}_l\PDDT{A}{\xi} +A|A|^2{\sf g}(\eta) + A|A|^4{\sf
h}(\eta),\hspace*{0.3cm}\mbox{in
LF}\\
\\
{\it \phi}_{23}=\D{{\it \phi}}_{23}=0,\hspace*{0.3cm}\mbox{on $\eta=-d^{-1}$}\\
\end{array}\right\}\no\\\label{nls22a}
\end{eqnarray}
where
\be
& &{\sf w}=-\zeta_{02}\psi_2'+\zeta_{22}\ts{\psi}_2'-2\ts{\zeta}_1\psi_3'-\psi_2''|\zeta_1|^2+\ts{\psi}_2''\zeta_1^2\no\\
& &{\mathscr W}=-\zeta_{02}{\it\Psi}_2'+\zeta_{22}\ts{\it\Psi}_2'-2\ts{\zeta}_1{\it\Psi}_3'-{\it\Psi}_2''|\zeta_1|^2+\ts{\it\Psi}_2''\zeta_1^2\no
\en
and
\be
{\sf h}(\eta)&=&\psi_3'(\ts{\psi}_2''-k_c^2\ts{\psi}_2)+\ts{\psi}_3(\ts{\psi}_2'''-k_c^2\ts{\psi}_2')-2\ts{\psi}_2'(\psi_3''-4k_c^2\psi_3)\no\\
&-&\ts{\psi}_2(\psi_3'''-4k_c^2\psi_3')-{\frak f}'(\psi_2''-k_c^2\psi_2)+{\frak f}'''\psi_2\no\\
{\mathscr H}(\eta)&=&{\it\Psi}_3'(\ts{\it\Psi}_2''-k_c^2\ts{\it\Psi}_2)+\ts{\it\Psi}_3(\ts{\it\Psi}_2'''-k_c^2\ts{\it\Psi}_2')-2\ts{\it\Psi}_2'({\it\Psi}_3''-4k_c^2{\it\Psi}_3)\no\\
&-&\ts{\it\Psi}_2({\it\Psi}_3'''-4k_c^2{\it\Psi}_3')-{\frak F}'({\it\Psi}_2''-k_c^2{\it\Psi}_2)+{\frak F}'''{\it\Psi}_2\no\\
\zeta_1&=&-\psi_1/\sigma_r=-{\it\Psi}_1/\sigma_r\no\\
\zeta_{02}&=&-[{\frak f}(0)+\ts{u}_{11}\zeta_1+\psi_1'\ts{\zeta}_1]/(u_b+a_1)\no\\
\zeta_{22}&=&-[\psi_2+\tf{1}{2}\zeta_1(u_{11}+\psi_1')]/\sigma_r\no\\
u_{11}&=&\psi_1'+u_b'\zeta_1\no\\
K_l&=&{\sf v}-(u_{02}\zeta_1-u_{22}\ts{\zeta}_1+2\ts{u}_{11}\zeta_{22})\no\\
K_u&=&{\mathscr V}-(u_{02}\zeta_1-u_{22}\ts{\zeta}_1+2\ts{u}_{11}\zeta_{22})\no\\
{\sf v}&=&-\zeta_{02}\psi_1'+\zeta_{22}\ts{\psi}_1'-2\ts{\zeta}_1\psi_2'-\psi_1''|\zeta_1|^2+\ts{\psi}_1''\zeta_1^2\no\\
{\mathscr V}&=&-\zeta_{02}{\it\Psi}_1'+\zeta_{22}\ts{\it\Psi}_1'-2\ts{\zeta}_1{\it\Psi}_2'-{\it\Psi}_1''|\zeta_1|^2+\ts{\it\Psi}_1''\zeta_1^2\no\\
u_{02}&=&{\frak f}'(0)+u_b'\zeta_{02}+(\zeta_1\ts{\psi}_1''+{\rm c.c.})+u_b''|\zeta_1|^2\no\\
u_{22}&=&\psi_2'+u_b'\zeta_{22}+\zeta_1\psi_1''+\tf{1}{2}u_b''\zeta_1^2\no\\
{\mathscr \alpha}_u &=& i{\it\Psi}_1\left[\DD{U}_b+2k_c^2(U_b-c_r)\right]
-\left[\frac{4k_c}{R_c}+i(U_b+a_1)\right]\left(\DD{\it\Psi}_1-k_c^2{\it\Psi}_1\right)\no\\
{\mathscr \alpha}_l &=& i\psi_1\left[\DD{u}_b+2k_c^2(u_b-c_r)\right]
-\left[\frac{4k_c}{\nu R_c}+i(u_b+a_1)\right]\left(\DD{\psi}_1-k_c^2\psi_1\right)\no\\
{\mathscr \beta}_u &=& i{\it\Psi}_{10}\left[\DD{U}_b+2k_c^2(U_b-c_r)\right]
-\left[\frac{4k_c}{R_c}+i(U_b+a_1)\right]\left(\DD{\it\Psi}_{10}-k_c^2{\it\Psi}_{10}\right)\nonumber\\
&+&a_2\left(\DD{\it\Psi}_1-k_c^2{\it\Psi}_1\right) +ik_c {\it\Psi}_1(3U_b-c_r+2a_1) -\frac{2}{R_c}
\left(\DD{\it\Psi}_1-3k_c^2{\it\Psi}_1\right)\no\\
{\mathscr \beta}_l &=& i\psi_{10}\left[\DD{u}_b+2k_c^2(u_b-c_r)\right]
-\left[\frac{4k_c}{\nu R_c}+i(u_b+a_1)\right]\left(\DD{\psi}_{10}-k_c^2\psi_{10}\right)\nonumber\\
&+&a_2\left(\DD{\psi}_1-k_c^2\psi_1\right) +ik_c \psi_1(3u_b-c_r+2a_1)
-\frac{2}{\nu R_c} \left(\DD{\psi}_1-3k_c^2\psi_1\right)\no\\
{\sf m}&=&ik_c(-{\frak f}'\psi_2'-\ts{\psi}_2'\psi_3'+2\psi_3\ts{\psi}_2''-\psi_3''\ts{\psi}_2)-ik_cK_l({\mathscr F}+k_c^2{\mathscr T})/\sigma_rR_c^2\no\\
&+&3Tik_c^5\zeta_1|\zeta_1|^2/2R_c^2+ik_c\rho{\sf p}_{13}\no\\
{\mathscr M}&=&ik_c(-{\frak f}'{\it\Psi}_2'-\ts{\it\Psi}_2'{\it\Psi}_3'+2{\it\Psi}_3\ts{\it\Psi}_2''-{\it\Psi}_3''\ts{\it\Psi}_2)-ik_cK_l{\mathscr F}/\sigma_rR_c^2\no\\
&+&3Tik_c^5\zeta_1|\zeta_1|^2/2R_c^2+ik_c\rho{\mathscr P}_{13}\no\\
{\sf p}_{13}&=&ik_c\rho\left[(\B{p}_{02}'-\B{p}_{22}')\zeta_1+(\zeta_{02}-\zeta_{22})\B{p}_{11}'+(|\zeta_1|^2-\tf{1}{2}\zeta_1^2)\B{p}_{11}''\right]\no\\
{\mathscr
P}_{13}&=&ik_c\rho\left[(\B{P}_{02}'-\B{P}_{22}')\zeta_1+(\zeta_{02}-\zeta_{22})\B{P}_{11}'+(|\zeta_1|^2-\tf{1}{2}\zeta_1^2)\B{P}_{11}''\right]\no
\en In (\ref{nls22}), the remaining coefficients are given by
Blennerhassett [2].

This system of equations has a unique solution provided the
inhomogeneous terms are orthogonal to the adjoint function.
Applying the solvability condition yields the amplitude equation
(\ref{1.1}).

The complex coefficients $\kappa, a_2$, $d_1$ and $\varpi$ in
equation (\ref{1.1}) are given by
\newpage
\be & &\kappa=\df{u_b'(\DD{\psi}+k_c^2\psi)}{\sigma_r\nu
R_c}-\df{\DDD{\psi}-3k_c^2\D{\psi}}{\nu
R_c}+ik_c\sigma_r\D{\psi}-\df{ik_c\psi(\mathscr{F}+k_c^2\mathscr{T})}{\sigma_rR_c^2}\no\\
& &a_2=-\df{\DD{\psi}{+k_c^2\psi}}{\nu
R_c}(u_b'-U_b')(a_1+c_r)\left(1-\df{u_b+a_1}{k_c\sigma_r}\right)\df{\phi}{k_c\sigma_r^2}+\df{\D{\psi}}{\nu
R_c}(1-\mu)(2k_c+1)\phi\no\\
& &\hspace*{0.3cm}+i\phi\df{k_c\mathscr{T}(3u_b-5c_r-2a_1)}
{\sigma_r^2R_c^2}\no\\
& &\hspace*{0.3cm}-\psi\left\{i\left[\phi
u_b'-\D{\phi}(u_b+a_1)\right]-i\rho\left[\Phi
U_b'-\D{\Phi}(U_b+a_1)\right]-\df{6k_c}{\nu
R_c}(\D{\phi}-\mu\D{\Phi})-\df{3}{\nu R_c}(\D{\phi}-\mu\D{\Phi})\right.\no\\
& &
\hspace*{0.3cm}\left.+i\phi\left[\df{\mathscr{F}(u_b-2c_r-a_1)(1-\rho)+k_c^2\mathscr{T}(3u_b-4c_r-a_1)}{\sigma_r^2
R_c^2}+(a_1+c_r)^2\df{\mathscr{F}(1-\rho)+k_c^2\mathscr{T}}{\sigma_r^3k_cR_c^2}\right]\right\}\no\\
& &a_1=\df{ic_r\phi}{\nu
k_cR_c\sigma_r^2}(u_b'-U_b')(\psi''+k_c^2\psi)-\df{2ik_c\phi\psi'(1-\mu)}{\nu
R_c}\no\\
& &\hspace*{0.3cm}-\psi\left[\phi u_b'-\phi'u_b-\rho(\Phi
U_b'-\Phi'U_b)+\df{6ik_c}{\nu R_c}(\phi'-\mu\Phi')\right.\no\\ & &
\hspace*{0.3cm}\left.+\phi\df{{\mathscr
F}(u_b-2c_r)(1-\rho)+k_c^2{\mathscr
T}(3u_b-4c_r)}{\sigma_r^2R_c^2}\right]\no\\
&
&d_1=\df{ik_c\psi}{R_c}\left[\df{\mathscr{F}(1-\rho)+k_c^2\mathscr{T}}{\sigma_rR_c^2}\phi+(1-\rho)
(\D{\phi}\sigma_r-\phi u_b')\right]\no\\
& &\hspace*{0.3cm}-\df{1}{\nu
R_c^2}\Int{-d^{-1}}{0}\psi(\phi^{iv}-2k_c^2\DD{\phi}+k_c^4\phi)\,d\eta\no\\
&
&\hspace*{0.3cm}-\df{\rho}{R_c^2}\Int{0}{1}\Psi(\Phi^{iv}-2k_c^2\DD{\Phi}+k_c^4\Phi)\,d\eta\no\\
 & &\varpi=\df{-i\phi}{\nu
k_cR_c\sigma_r^2}(u'_b-U'_b)(\psi''+k_c\psi)+\psi\left[\rho\Phi'-\phi'-\phi\df{{\mathscr
F}(1-\rho)+k_c^2{\mathscr T}}{\sigma_r^2R_c^2}\right]\no\\
&
&\hspace*{0.3cm}+\Int{-d^{-1}}{0}\psi(\phi''-k_c^2\phi)\,d\eta+\rho\Int{0}{1}\Psi(\Phi''-k_c^2\Phi)\,d\eta\no
\en where $\Phi, \phi$ satisfies the Orr-Sommerfeld equation for
the basic flow with $\Psi, \psi$ satisfying their adjoints.

It is important to note that in the formulation presented here quantities such as $\Phi$ and $\phi$,
and their adjoints, are functions of $\eta$
alone, but, by virtue of (\ref{2.6}) they are functions of both spacial coordinates $x$, $y$ and time $t$.
Also, since the Jacobian of the transformation is $1+O(ak)$, the resulting equations expressed
in curvilinear coordinates are, to first order in $ak$, symbollically identical to their
Caretesian counterpart. Thus, the solutions for the basic flow, in curvilinear coordinates, are the same as those given by
(\ref{2.4}) and (\ref{6.1}), in Cartesian coordinates, with $y$ simply replaced with $\eta$. The error in this approximation
is $O(ak)^2$ and may be neglected since $ak\ll 1$.
\section{Numerical Schemes}\label{S:3}
The complex cubic--quintic nonlinear Schr\"odinger equation
(\ref{1.1}) is solved numerically, using a second-order central
difference
for the spacial part and 
the third-order explicit Runge-Kutta method for the transient
part, whose complex coefficients are determined from
the system of ordinary differential equations (\ref{nls14}), (\ref{nls18}) and (\ref{nls18}) which coupled via
interface conditions  which are solved numerically
using the finite difference scheme as described below. First, we
rewrite the system in a compact form as
\begin{eqnarray}
\left.\begin{array}{lcl}
\Phi=\Phi' = 0 & & {\rm on} \,\, \eta=1,\\
\\
\Phi^{iv}+C(\eta)\Phi''+D(\eta)\Phi + 2E(\eta)\Phi' = F(\eta)
& & {\rm in} \,\, 0<\eta<1,\\
\\
\sum_{p=0}^3 FL_{np} \phi^{(p)}+FLNH_n =
\sum_{p=0}^3 FU_{np} \Phi^{(p)}+FUNH_n, & & \eta=0,\,n=1,2,3,4,\\
\\
\phi^{iv}+c(\eta)\phi''+d(\eta)\phi + 2e(\eta)\phi' = f(\eta)
& & {\rm in} \,\, -d^{-1}<\eta<0,\\
\\
\phi=\phi' = 0 & & {\rm on} \,\, \eta=-d^{-1},\\
\end{array}\right\}\no\\\label{numeric}
\end{eqnarray}
where $\Phi=({\it\Phi}_{11},{\it\Phi}_{22},\Psi)$, $\phi=(\phi_{11},\phi_{22},\psi)$ and the definitions of
the coefficients $c$, $C$, $d$, $D$, $e$, $E$, $FL_{np}$,
$FU_{np}$, $FLNH_n$, $FUNH_n$, $f$, $F$ are those given by system
of equation (\ref{nls14}), (\ref{nls18}) and (\ref{adjor}), respectively.

We approximate
\begin{eqnarray}
\left.\begin{array}{lcccl}
\Phi=\left(1+\frac16 \delta^2 - \frac{1}{720}\delta^4\right)G & & & & \mbox{for the upper fluid}\\
\\
\phi=\left(1+\frac16 \delta^2 - \frac{1}{720}\delta^4\right) g & & & & \mbox{for the lower fluid}\\
\end{array}\right\}\label{fd1}
\end{eqnarray}
where $\delta$ is the usual central difference operator, i.e.,
$\delta g_i = g_{i+1/2}-g_{i-1/2}$. Then
\begin{eqnarray}
h^4\phi^{iv} &=& \delta^4 g + O\left(\delta^{10}\right)\label{eq1} \\
h^2\phi'' &=& (\delta^2+\frac{1}{12}\delta^4) g +O\left(\delta^{6}\right) \label{eq2}\\
h\phi' &=& \mu \delta g + O \left(\delta^{5}\right) \label{eq3} 
\end{eqnarray}
Here $h = \eta_i-\eta_{i-1}$ and $\mu$ 
is defined as $\mu g_i = \frac12\left(g_{i+1/2}+g_{i-1/2}\right)$.

For $\phi'''$ we use the following approximation
\begin{eqnarray}\label{fd2}
&h^3\phi'''=\mu\delta^3 \left(1+\frac16 \delta^2 -
\frac{1}{720}\delta^4\right)^{-1} \phi -\frac{1}{12}\mu
\delta\left(h^4\phi^{(iv)}\right) + O\left(\delta^7\right). &
\end{eqnarray}
Using (\ref{fd2}) and the appropriate equation of motion from (\ref{numeric}),
we have
\begin{eqnarray}\label{fd3}
&h^3\phi'''=\mu\delta^3 g +\frac{1}{12}h^2\mu \delta
\left(c\delta^2 g + h^2 d g +2he\mu \delta g -h^2 f \right) +
O\left(\delta^7\right). &
\end{eqnarray}
Hence we maintain $O\left(h^4\right)$ accuracy and a five point
formula for the third derivative in the interface. The similar
expressions result for the derivatives of $\Phi$. Substituting
(\ref{eq1})--(\ref{eq3}) in (\ref{numeric}), we obtain the
following finite difference approximation.
\begin{eqnarray}\label{fd3}
&{\mathscr A}_1 g_{i-2} + {\mathscr A}_2 g_{i-1} + {\mathscr A}_3
g_{i} + {\mathscr A}_4 g_{i+1} + {\mathscr A}_5 g_{i+2} =h^4 f_i,
&
\end{eqnarray}
where
\begin{eqnarray}
&{\mathscr A}_1 = {\mathscr A}_5  = 1+\frac{1}{12}h^2c_i & \\
&{\mathscr A}_2 = -4+\frac23h^2c_i + \frac16 h^4d_i -h^3e_i &\\
&{\mathscr A}_3 = 6-\frac32h^2c_i + \frac23 h^4d_i &\\
&{\mathscr A}_4  = {\mathscr A}_2 + 2h^3e_i &
\end{eqnarray}
where the subscript $i$ implies evaluation at $\eta=\eta_i$.
Similar expressions result for the approximation to $\Phi$. Thus,
the finite difference approximation to the coupled ODEs system
with the boundary conditions and the interface condition
(\ref{numeric}) leads to the matrix equation ${\bf{\mathscr A} x}
= {\bf b}$ with the following form: {\tiny
\[
\left(
\begin{array}{lcccccccccccccccccccccr}
\ * & * & * &   &   &   &   &   &   &   &   &   &   &   &   &   &   &   &   &   &   &   &  \\
\ * & * & * & * &   &   &   &   &   &   &   &   &   &   &   &   &   &   &   &   &   &   & \\
\ * & * & * & * & * &   &   &   &   &   &   &   &   &   &   &   &   &   &   &   &   &   & \\
\   & * & * & * & * & * &   &   &   &   &   &   &   &   &   &   &   &   &   &   &   &   & \\
\   &   &   &   &\vdots &\vdots &   &   &   &   &   &   &   &   &   &   &   &   &   &   & \\
\   &   &   &   &   & * & * & * & * & * &   &   &   &   &   &   &   &   &   &   &   &   & \\
\   &   &   &   &   &   & * & * & * & * & * &   &   &   &   &   &   &   &   &   &   &   & \\
\   &   &   &   &   &   &   & * & * & * & * & * &   &   &   &   &   &   &   &   &   &   & \\
\   &   &   &   &   &   &   & * & * & * & * & * & * & * & * & * & * &   &   &   &   &   & \\
\   &   &   &   &   &   &   & * & * & * & * & * & * & * & * & * & * &   &   &   &   &   & \\
\   &   &   &   &   &   &   & * & * & * & * & * & * & * & * & * & * &   &   &   &   &   & \\
\   &   &   &   &   &   &   & * & * & * & * & * & * & * & * & * & * &   &   &   &   &   & \\
\   &   &   &   &   &   &   &   &   &   &   &   & * & * & * & * & * &   &   &   &   &   & \\
\   &   &   &   &   &   &   &   &   &   &   &   &   & * & * & * & * & * &   &   &   &   & \\
\   &   &   &   &   &   &   &   &   &   &   &   &   &   & * & * & * & * & * &   &   &   & \\
\   &   &   &   &   &   &   &   &   &   &   &   &   &   &   &   &   &\vdots &\vdots &   & \\
\   &   &   &   &   &   &   &   &   &   &   &   &   &   &   &   &   & * & * & * & * & * & \\
\   &   &   &   &   &   &   &   &   &   &   &   &   &   &   &   &   &   & * & * & * & * & * \\
\   &   &   &   &   &   &   &   &   &   &   &   &   &   &   &   &   &   &   & * & * & * & * \\
\   &   &   &   &   &   &   &   &   &   &   &   &   &   &   &   &   &   &   &   & * & * & * \\
\end{array}
\right) \left(
\begin{array}{c}
G_{NU} \\
\cdot \\
\cdot \\
\cdot \\
\cdot \\
G_2 \\
G_1 \\
G_0\\
G_{-1}\\
G_{-2}\\
G_{-3}\\
g_{-3} \\
g_{-2} \\
g_{-1} \\
g_0 \\
g_1 \\
g_2\\
\cdot \\
\cdot \\
\cdot \\
\cdot \\
g_{NL}
\end{array}
\right) = \left(
\begin{array}{c}
F_{NU} \\
\cdot \\
\cdot \\
\cdot \\
\cdot \\
F_2 \\
F_1 \\
F_0\\
F_{-1}\\
F_{-2}\\
F_{-3}\\
f_{-3} \\
f_{-2} \\
f_{-1} \\
f_0 \\
f_1 \\
f_2\\
\cdot \\
\cdot \\
\cdot \\
\cdot \\
f_{NL}
\end{array}
\right)
\]
} where the entries of the matrix ${\bf \mathscr A}$ and unknown
vector ${\bf x}$ are complex numbers.

\subsection{Complex cubic--quintic nonlinear Schr\"odinger equation}
Let $h=(b-a)/n$ be the grid spacing and $x_j=a+jh, \,
j=0,1,\cdots,n$ be the grid points. Define $A_j(t)$ as an
approximation to $A(x_j,t),\,j=0,1,\cdots,n$, ${\bf A}(t)=
(A_1(t),\cdots,A_{n-1}(t))^{\sf T}$ and ${\bf A}_0(x)=
(A_0(x_1),\cdots,A_{0}(x_{n-1}))^{\sf T}$. Let $F(A_j(t))=
\df{d_1}{d_{1r}}A_j(t)+\kappa A_j(t)|A_j(t)|^2+\varpi
A_j(t)|A_j(t)|^4$. Using the central difference approximation to
$\partial_x^2 A(x_j,t)$, we write the semi--discretization of the
IBVP for equation (\ref{1.1}) as the following system of ODEs
\begin{eqnarray}
&\dot{\bf A}(t) = {\mathscr B}{\bf A}(t) + {\bf F}({\bf A}(t)), \,
\forall t>0,
& \label{eq:4}\\
&{\bf A}(0)={\bf A}_0, & \label{eq:5}
\end{eqnarray}
where ${\bf F}({\bf
A}(t))=\left(F(A_1(t)),\,\cdots,\,F(A_n(t))\right)^{\sf T}$. The upper
dot in (\ref{eq:4}) indicates derivative with respect to $t$ and
${\mathscr B}$ is the finite difference matrix.

To construct an integration scheme to solve the ODE system
(\ref{eq:4})--(\ref{eq:5}), let  $t_{n+1} = t_n +\Delta t$, and
let ${\bf A}^{n}$ denote the value of the variable ${\bf A}$ at
time $t_n$. Employing a low storage variant third--order
Runge--Kutta scheme [12], we write the fully discrete system as
\begin{equation}\label{RK3}
\begin{array}{lll}
&{\bf Q}_1 = \Delta t {\bf G}({\bf A}^n), &
{\bf A}_1 = {\bf A}^n + \frac13 {\bf Q}_1, \\
\\
&{\bf Q}_2 = \Delta t {\bf G}({\bf A}_1)- \frac59{\bf Q}_1, &
  {\bf A}_2 = {\bf A}_1 + \frac{15}{16} {\bf Q}_2,\\ 
  \\
&{\bf Q}_3 = \Delta t {\bf G}({\bf A}_2)-\frac{153}{128}{\bf Q}_2,
&
{\bf A}^{n+1} = {\bf A}_2 + \frac{8}{15} {\bf Q}_3, \\ 
\end{array}
\end{equation}
where ${\bf G}({\bf A}^n)={\mathscr B}{\bf A}^n + {\bf F}({\bf A}^n)$.

The numerical scheme is parallelized for distributed memory
clusters of processors or heterogeneous networked computers using
the MPI (Message Passing Interface) library and implemented in
FORTRAN.
\section{Results}\label{S:4}
For the numerical simulation reported in this paper, we choose
${\mathscr F}=10^3k_s^3$ and ${\mathscr T}=10^3(1-\rho)k_s$ for
the present case of wind blowing over waves. Here the quantity
$k_s$ is defined as \be k_s=D_u\sqrt{\df{(\rho_w-\rho_a)g}{S}}\no
\en with $\rho_a=1.225\times 10^{-3}\,{\rm g/cm^3}$,
$\rho_w=0.9991\,{\rm g/cm^3}$, $g=981\,{\rm cm/s^2}$,
$S=73.5\,{\rm dyn/cm}$, and we choose $D_\ell=D_u$ such that
$d=1$. The kinematic viscosities are taken to be
$\nu_a=0.145\,{\rm cm^2/s}$ and $\nu_w=1.138\times 10^{-2}\,{\rm
cm^2/s}$.

The velocity profiles for the basic flow, for the case of
turbulent boundary layer, are those given by (\ref{2.1}). The
velocity profiles for pPf and pCf, for the upper fluid, satisfy
\begin{eqnarray}
\left.\begin{array}{lll}
U_b=U_1, & & \mbox{on}\hspace*{0.25cm} y=D_u\\
\df{G}{\rho_a}+\nu_w\ODDT{U_b}{y}=0,
& & \mbox{in upper fluid}\\
u_b=U_b,\hspace*{0.25cm}\mu_w\df{du_b}{dy}=\mu_a\df{dU_b}{dy},
& & \mbox{on}\hspace*{0.25cm} y=0\\
\df{G}{\rho_w}+\nu_w\ODDT{u_b}{y}=0, & & \mbox{in lower fluid}\\
u_b=0, & & \mbox{on}\hspace*{0.25cm} y=-D_\ell\\
\end{array}\right\}\label{6.1}
\end{eqnarray}
The solution of this system is given by [2] \be
& &U_b(y)=-{\sf A}(y-1)\left[y+\df{\mu(1+d)}{d(\mu+d)}\right]+{\sf B}(yd+\mu)\no\\
\no\\
& &u_b(u)=-\df{\mu}{2}{\sf A}(yd+1)\left(y-\df{1+d}{\mu+d}\right)+\mu{\sf B}(yd+1)\no
\en
where
\be
& &{\sf A}=U_p\left[\df{\mu+d^2}{6d^2(1+d)}+\df{1}{2}\df{\mu(1+d)}{d(\mu+d)}\right]^{-1}\no\\
\no\\
& &{\sf B}=(1-U_p)\df{2(1+d)}{\mu+d(d+2\mu)}\no
\en
and
\be
U_p=\df{2(1+d)d}{\mu+d(d+2\mu)}\left[\df{\mu+d^2}{6d^2(1+d)}+\df{\mu}{2d}\df{1+d}{\mu+d}\right]\left\{1+\df{\mu+d^2}{3d[\mu+d(d+2\mu)]}\right.& &\no\\
\left.+\df{\mu(1+d)}{d(\mu+d)}\left[1+\df{d(1+d)}{\mu+d(d+2\mu)}\right]\right\}^{-1}\no
\en
Note that, for pPf $U_p=1$ which implies that he upper boundary is stationary, and the flow is
driven by the pressure gradient. In contrast, for pCf $U_p=0$ and the motion is generated soley
by the relative motion of the boundaries.

Computations were performed on a Linux cluster (zeus.db.erau.edu:
256 Intel Xeon 3.2GHz 1024 KB cache 4GB with Myrinet MX, GNU
Linux) at Embry-Riddle University. In all simulations presented
here, we use the spacial domain of $0\le \xi \le 3$ and temporal
domain $0\le \tau\le 15$, the typical values of the parameters
$a=0.3$, $l=0.5$ for the amplitude and the wavelength of the
Stokes wave profile, respectively, and the other parameters of the
model (\ref{1.1}) are given in Table 1 below.
\begin{table}
\begin{center}
{\tiny 
\begin{tabular}{|c|cccc|}
\hline Profile & $d_1$ & $a_2$ & $\kappa$ & $\varpi$ \\
\hline pCf & $(0.895\times 10^{-5}, 0.217\times 10^{-2})$ &
$(0.895\times 10^{-5}, 0.217\times 10^{-2})$ &
$(-29.7, 237)$ & $(-5.94, 47.4)$ \\
\hline pPf & $(0.404\times 10^{-5}, 0.742\times 10^{-3})$ &
$(0.557\times 10^{-3}, -0.17\times 10^{-1})$ &
$(-9.95, 912.3)$ & $(-2.49, 228)$ \\
\hline LBL & $(0.554\times 10^{-5}, 0.275\times 10^{-2})$ &
$(0.165\times 10^{-2}, -0.159)$ &
$(-0.221, 20.61)$ & $(-0.0737, 6.87)$ \\
\hline TBL & $(0.222\times 10^{-2}, 0.376\times 10^{-3})$ &
$(0.042\times 10^{-3}, 0.11\times 10^{-1})$ &
$(-19.2, 1142)$ & $(-9.6, 571)$ \\
\hline
\end{tabular}
}
\begin{center}
\caption{\footnotesize Parameters for various models}
\end{center}
\end{center}
\end{table}

\begin{figure}
\vspace{1pc}
    \begin{center}
        \includegraphics[width=4.5cm]{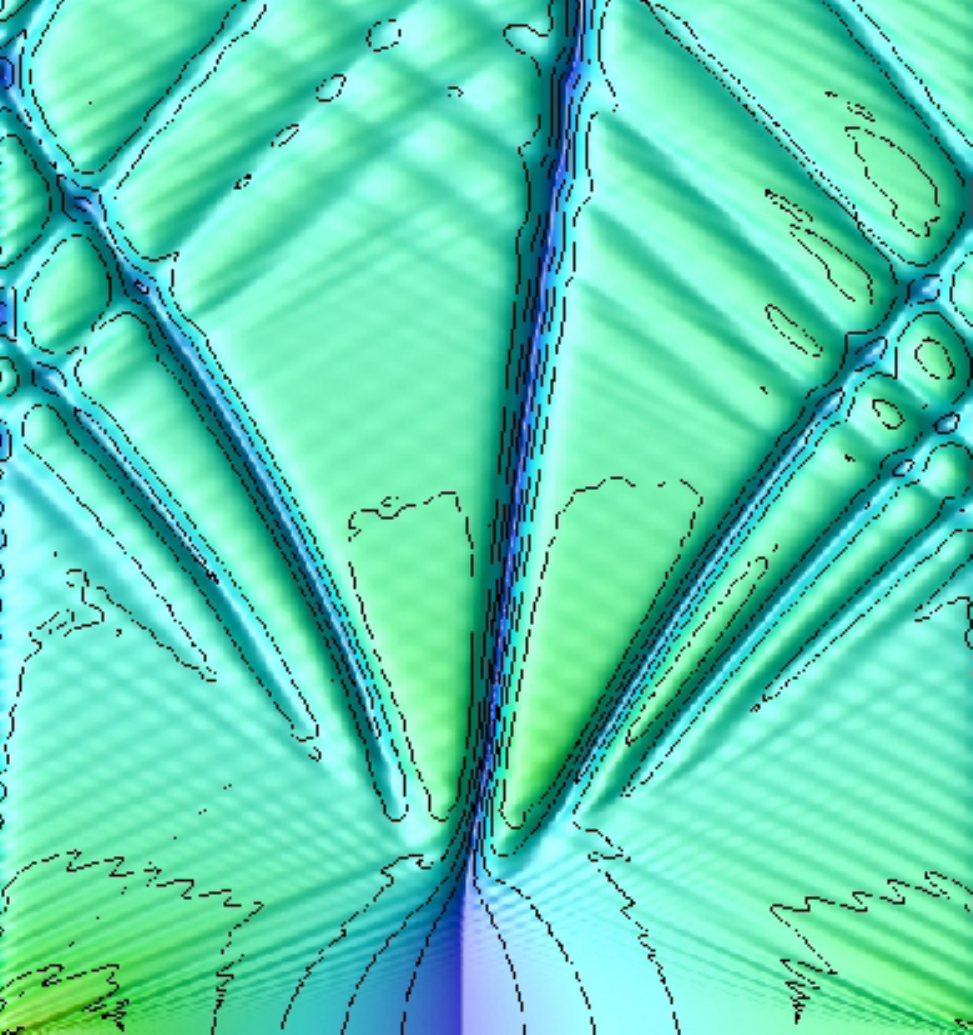}
\hspace*{1cm}
        \includegraphics[width=4.5cm]{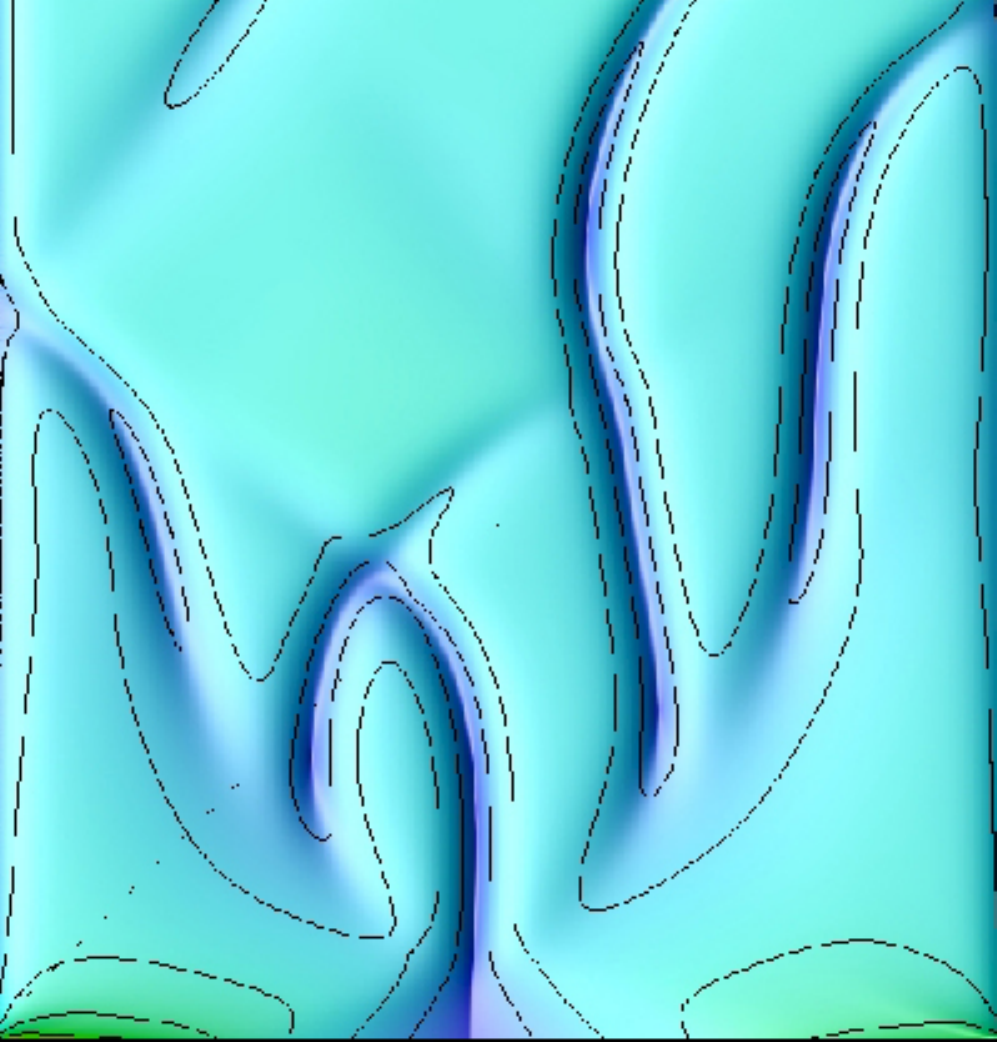}
    \end{center}
    \caption[oblique wave]{\footnotesize
Left: the surface profile for pCf in the air; Right: the surface
profile for pPf in the air.
} \label{fig:Fig1}
\end{figure}
\begin{figure}
\vspace{1pc}
    \begin{center}
        \includegraphics[width=4.5cm]{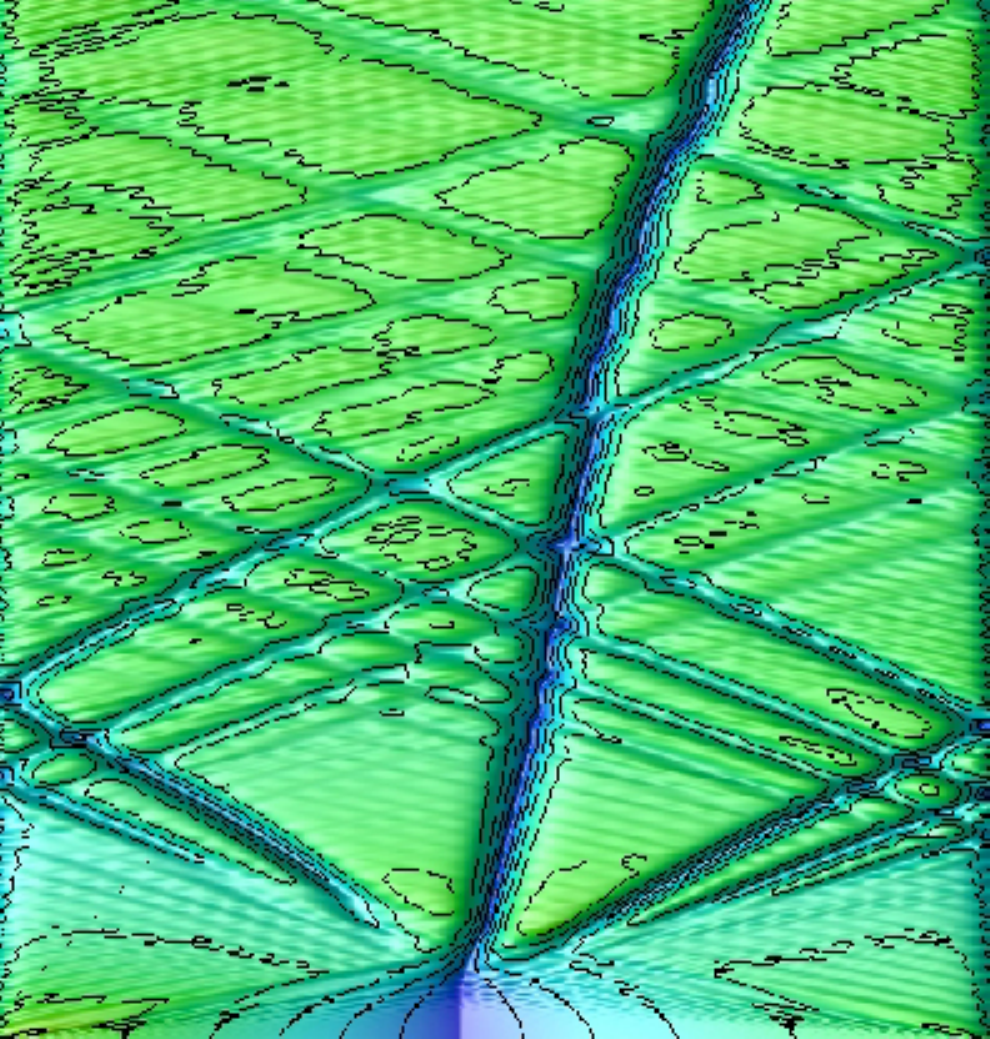}
    \hspace*{1cm}
        \includegraphics[width=4.5cm]{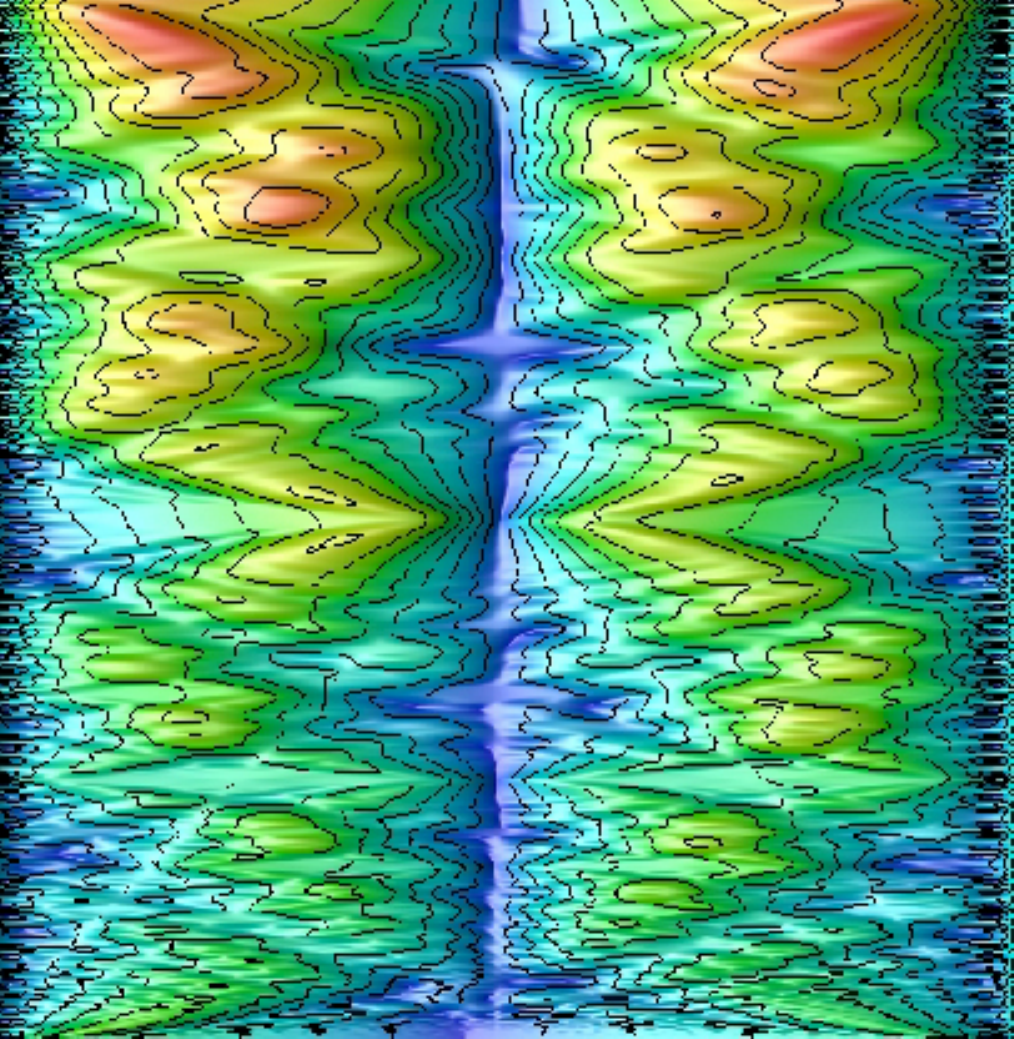}
    \end{center}
    \caption[oblique wave]{\footnotesize
Left: the surface profile for LBL in
the air, Right: the surface profile for TBL in the air.} \label{fig:Fig1}
\end{figure}
\begin{figure}
\vspace{1pc}
    \begin{center}
        \includegraphics[width=4.5cm]{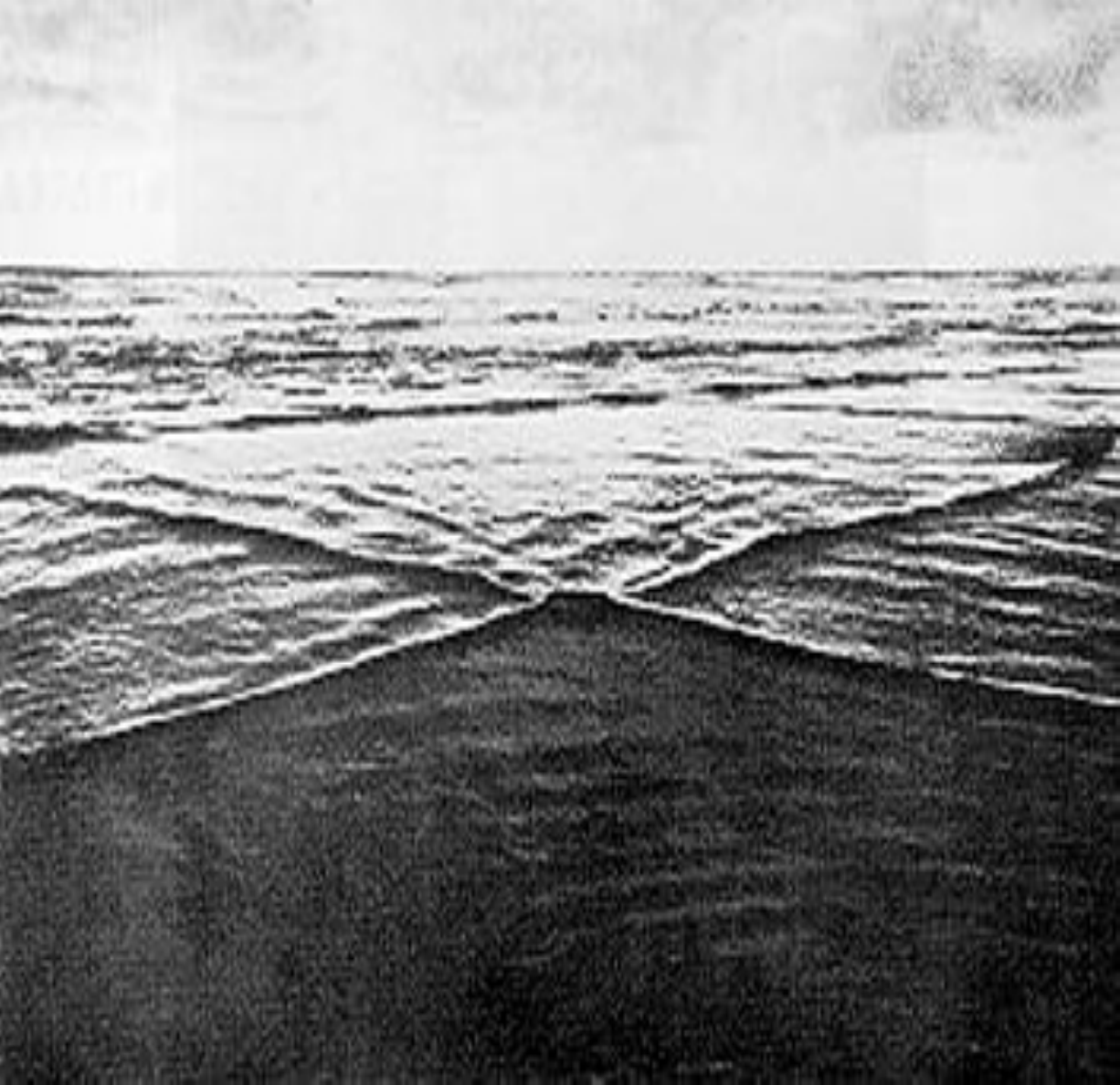}
    \hspace*{1cm}
\includegraphics[width=4.5cm]{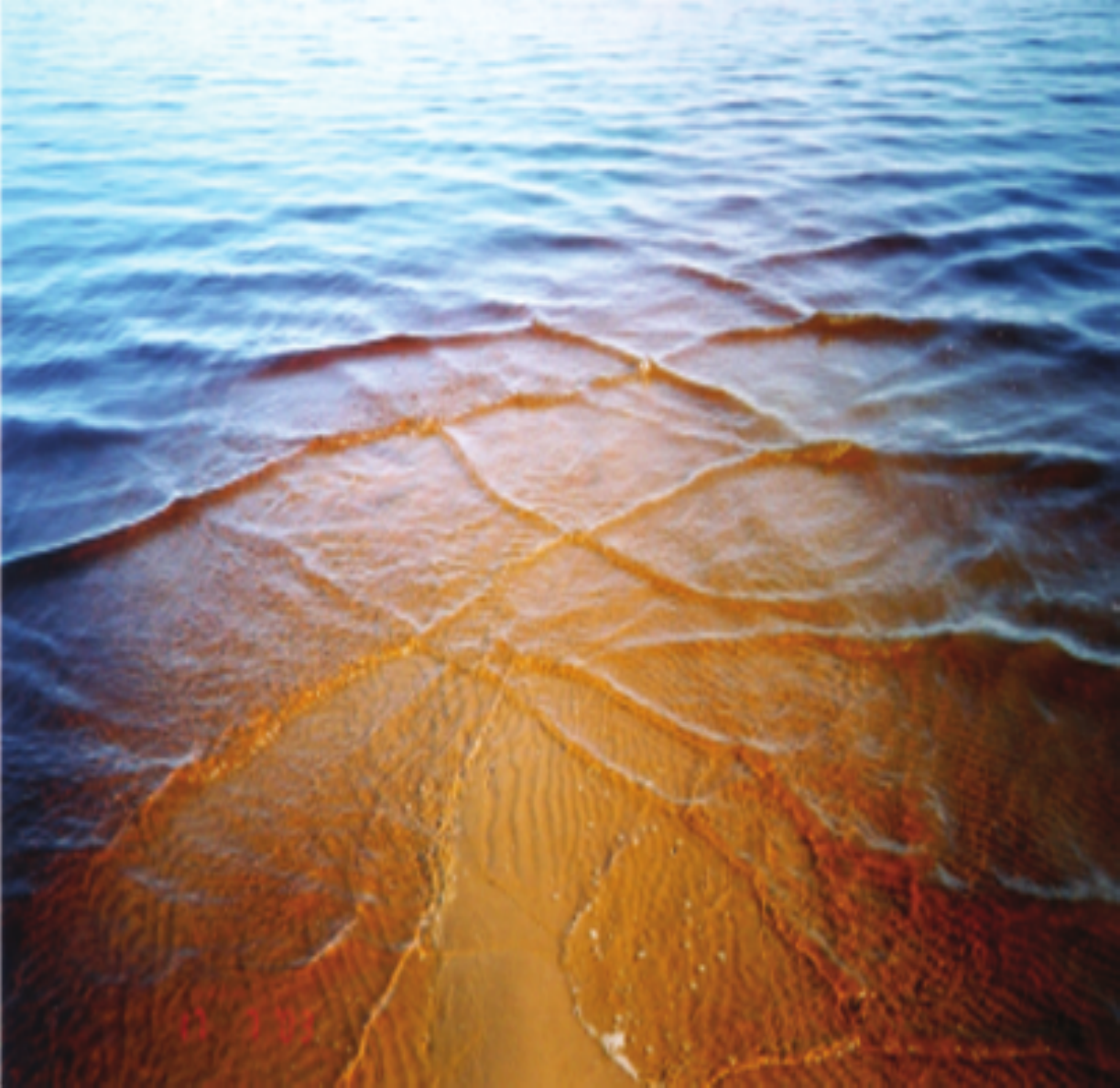}
    \end{center}
    \caption[oblique wave]{\footnotesize
Wind generated solitons in shallow water.}\label{fig:Fig2}
\end{figure}

Figure 2 shows the surface structure for pCf (left) and pPf
(right). For the case of pPf we observe the classical `snake' type
solitons [9], whereas for pCf we see a different structure similar
to colliding solitons which, in time, bifurcates to weaker soliton
structure.

In figure 3 the surface structure for LBL (left) and TBL (right)
are depicted. As can be seen, the soliton structures for TBL
resembles the chaotic solitons [9]. This is, of course, not
surprising as turbulence generates more complex surface agitation
compared to its laminar counterpart. We may term this as a violent
surface motion.

The most intriguing results obtained is that shown in figure 3 for
LBL. This results clearly confirms the generation of solitons by
wind, which has commonly been observed is shallow waters, c.f.
figure 4.

\section{Conclusions}

In this paper we presented an asymptotic theory for linear
stability of non-linear air-sea interfaces by taking into account
the motion in the air and the water. The two motions are then
matched at the free surface via the usual kinematic, dynamic and
free stress boundary conditions. The flow above water is assumed
to be turbulent, with a mean logarithmic velocity profile. On the
other hand, a plane Poiseuille flow is assumed for the motion in
the water. Also presented, is the linear stability calculations of
two fluid interface which suggest the flow become unstable for
wavenumbers above that given by equation (23). Finally, the growth
of surface waves by the shear wind blowing over is calculated. It
is shown and extra term arises in the expression for the
dimensionless energy transfer parameter which is due to the
coupling between wind and surface waves.

In the second part of this paper [3], we use a non-linear
hydrodynamic stability [7] to derive the quintic non-linear
Schr\"odinger equation (1) with complex coefficients. We then use
the above analysis to determine the complex amplitude of the
equation (1).

Using a nonlinear stability analysis of a coupled air-sea system
is performed. An amplitude equation, namely the quintic nonlinear
Schr\"odinger equation with complex coefficients is derived. The
complex coefficients are obtained from the linear stability of the
same system in the neighborhood of the critical Reynolds number.
Four different profiles for the shear flow for the airflow over
the interface is considered, namely, plane Couette flow (pCf),
plane Poiseuille flow (pPf), laminar and turbulent boundary layer
(L,TBL) profiles. For each of the above cases the shear flow
counterpart in the water is assumed to be a pPf.

It is shown that the above amplitude equation produces `snake'
solitons for pCf, pPf and LBL profiles, with striking
similarities. On the other hand, for TBL we observe a very violent
surface motion. For cases of pCf and LBL remarkable similarity is
observed with observations made at sea.

We conclude that the effect of nonlinearity in the airflow over
the sea surface is much larger than nonlinear interactions in the
water, and hence it is not possible to decouple the motion in the
air and the water for finite amplitude wind-wave interactions. We
remark, under the conditions shown in this paper, the generation
of solitons by wind is possible particularly in shallow waters.

\end{document}